\documentclass[a4paper]{revtex4}
\usepackage[english]{babel}
\usepackage{array,booktabs}
\usepackage{array} 
\usepackage{lipsum}   
\usepackage{calc}
\usepackage{pdflscape}
\usepackage{color}
\usepackage{float}
\usepackage[font=small,labelfont=bf]{caption}
\usepackage{graphicx}
\usepackage{amsmath}
\usepackage[nodisplayskipstretch]{setspace}

\usepackage{setspace}
\usepackage{tabularx}

\usepackage{float}
\usepackage{color}
\usepackage{amsmath}
\usepackage{float}
\usepackage{calc}
\usepackage{pdflscape}
\usepackage{color}
\usepackage{float}
\usepackage{subfigure}
\usepackage[font=small,labelfont=bf]{caption}
\usepackage{graphicx}
\begin{document}{\setlength\abovedisplayskip{4pt}}

\title{Octant Degeneracy, Quadrant of CPV phase at Long Baseline $\nu$ Experiments and Baryogenesis}
\author{Kalpana Bora}
\affiliation{Department Of Physics, Gauhati University, Guwahati-781014, India}
\author{Gayatri Ghosh}
\affiliation{Department Of Physics, Gauhati University, Guwahati-781014, India}
\author{Debajyoti Dutta}
\affiliation{Harish-Chandra Research Institute, Chhatnag Road, Jhunsi, Allahabad 211019, India}


\begin{abstract}
In a recent work by two of us, we have studied, how CP violation discovery potential can be improved at long baseline neutrino experiments (LBNE/DUNE), by combining with its ND (near detector) and reactor experiments. In this work, we discuss how this study can be further analysed to resolve entanglement of the quadrant of leptonic CPV phase and Octant of atmospheric mixing angle $ \theta_{23} $, at LBNEs. The study is done for both NH (Normal hierarchy) and IH (Inverted hierarchy), HO (Higher Octant) and LO (Lower Octant). We show how leptogenesis can enhance the effect of resolving this entanglement, and how possible values of the leptonic CPV phase can be predicted
in this context. Carrying out numerical analysis based on the recent updated experimental results for neutrino mixing angles, we predict the values of the leptonic CPV phase for 152 possible cases. We also confront our predictions of the leptonic CPV phase with the updated global fit and find that five values of $ \delta_{CP} $ are favoured by BAU constraints. One of the five values matches with the recent global fit value of  $ \delta_{CP} $ (leptonic CPV phase) close to 1.41$ \pi $  in our model independent scenario.  A detailed analytic and numerical study of baryogenesis through leptogenesis is performed in this framework in a model independent way. 
\end{abstract}

\maketitle
\section{Introduction}	
\setlength{\baselineskip}{13pt}
Today, physics is going through precision era-this is more so for Neutrino physics. With the measurement of reactor angle $\theta_{13}$ \cite{Fgli,Frero,Garc} precisely by reactor experiments, the unknown quantities left to be measured in neutrino sector are $-$ leptonic CP violating phase \cite{DK,MG,PT,Kang,LHCb,Patrik}, octant of atmospheric angle $\theta_{23}$ \cite{KD,Gonzalez,Animesh,Choubey,Daljeet}, mass hierarchy, nature of neutrino etc. Long baseline neutrino experiments (LBNE \cite{LBNE,Akiri},  NO$ \nu $A \cite{Ayres} , T2K \cite{T2K}, MINOS \cite{Minos}, LBNO \cite{DA} etc) may be very promising, in measuring many of these sensitive parameters.
\par 
Exploring leptonic CP violation (CPV) is one of the most demanding tasks in future
neutrino experiments \cite{Branco}. The relatively large value of the reactor mixing angle $ \theta_{13} $ measured with a high precision in neutrino experiments \cite{F.P} has opened up a wide range of possibilities to examine CP violation in the lepton sector. The leptonic CPV phase can be induced by the PMNS
neutrino mixing matrix \cite{Pcarvo} which holds, in addition to the three mixing angles, a Dirac type CP
violating phase in general as it exists in the quark sector, and two extra phases if neutrinos are Majorana particles. Even if we do not yet have significant evidence for leptonic CPV, the current global fit to available neutrino data manifests nontrivial values of the Dirac-type CP phase \cite{Capp,Gon}. In this context, possible size of leptonic CP violation detectable through neutrino oscillations can be predicted. Recently, \cite{DK}, two of us have explored possibiities of improving CP violation discovery potential of newly planned Long-Baseline Neutrino Experiments (earlier LBNE, now called DUNE) in USA. In neutrino oscillation probability expression P($ \nu_{\mu}\rightarrow \nu_{e}$) relevant for LBNEs, the term due to significant matter effect, changes sign when oscillation is changed from neutrino to antineutrino mode, or vice-versa. Therefore in presence of matter effects, CPV effect is entangled and hence, one has two degenerate solutions - one due to CPV phase and another due to its entangled value. It has been suggested to resolve this issue by combining two experiments with different baselines \cite{Varger,Minakata}. But CPV phase measurements depends on value of reactor angle $\theta_{13}$, and hence precise measurement of $\theta_{13} $ plays crucial role in its CPV measurements. This fact was utilised recently by two of us \cite{DK}, where we explored different possibilities of improving CPV sensitivity for LBNE, USA. We did so by considering LBNE with \\
1. Its ND (near detector).\\
2. And reactor experiments.
\par 
We considered both appearance ($ \nu_{\mu}\rightarrow \nu_{e}$) and disappearance ($ \nu_{\mu}\rightarrow \nu_{e}$) channels in both neutrino and antineutrino modes. Some of the observations made in \cite{DK} are\\
1. CPV discovery potential of LBNE increases significantly when combined with near detector and reactor experiments.\\ 
2. CPV violation sensitivity is more in LO (lower octant) of atmospheric angle $\theta_{23}$, for any assumed true hierarchy.\\
3. CPV sensitivity increases with mass of FD (far detector).\\
4. When NH is true hierarchy, adding data from reactors to LBNE improve its CPV sensitivity irrespective of octant.
\par 
Aim of this work is to critically analyse the results presented in \cite{DK}, in context of entanglement of quadrant of CPV phase and octant of $\theta_{23} $, and hence study the role of leptogenesis (and baryogenesis) in resolving this enganglement. Though in \cite{DK},
 we studied effect of both ND and reactor experiments on CPV sensitivity of the LBNEs, in this work we have considered only the effect
 of ND. But similar studies can also be done for the effect of Reactor experiments on LBNEs as well. The details of LBNE and ND are same
 as in \cite{DK}. Following the results of \cite{DK}, either of the two octants is favoured, and the enhancement of CPV sensitivity with
respect to its quadrant is utilized here to calculate the values of lepton-antilepton symmetry. This is done considering two cases of the
rotation matrix for the fermions - CKM only, and CKM+PMNS. Then, this is used to calculate the 
value of BAU. This is an era of precision measurements in neutrino physics. We therefore consider variation of $\Delta m^{2}_{31}$ 
within its 1$\sigma$, 2$\sigma$ and 3$\sigma$ range values. We calculate baryon to photon ratio, and compare with its experimentally 
known best fit value. We observe that the BAU can be explained most favourably for five possible cases explored here: IH, $\delta_{CP}= 1.43 \pi$ and HO of $ \theta_{23} $; IH, $\delta_{CP}= 0.5277 \pi$ and HO of $ \theta_{23} $; IH, $\delta_{CP}= 0.488 \pi$ and LO of $ \theta_{23} $; IH, $\delta_{CP}= 0.383 \pi$ and HO of $ \theta_{23} $; IH, $\delta_{CP}= 1.727 \pi$ and LO of $ \theta_{23} $. It is worth mentioning that the value of $\delta_{CP} = 1.43 \pi$ favoured by our calculation here is close to the central value of $ \delta_{CP} $ from the recent global fit result \cite{Gon, kol}. We also find that for variation of $\Delta m^{2}_{31}$, within its 1$\sigma$ range, the calculated values of $\eta_B$ for all possible five cases mentioned above lie in the allowed range of its best fit value. But for 3 $ \sigma $  variation of $\Delta m^{2}_{31}$, some of its values at its  3$ \sigma $ C.L are disfavoured. Also for the  variation of $\theta_{13}$ within its 3 $ \sigma $ C.L, its values around 9.0974 are favoured, as far as matching with the best fit values of $\eta_B$  are concerned. These results could be important keeping in view that the quadrant of leptonic CPV phase, and octant of atmospheric mixing angle $\theta_{23}$ are yet not fixed. Also, they are significant in context of precision measurements on neutrino oscillation parameters. 

\par
The paper is organized as follows. In Section II, we discuss entanglement of quadrant of  CPV phase and octant of $ \theta_{23} $. In Section III, we present a review on leptogenesis and baryogenesis. In Sec. IV we show how the  
baryon asymmetry (BAU) within the SO(10) model, by using two distinct forms for the lepton CP asymmetry, can be used to break the entanglement. Sec. V summarizes the work.

\section{CPV Phase and Octant of $\theta_{23}$}

As discussed above, from Fig. 3 of \cite{DK}, we find that by combining with ND and reactor experiments, CPV sensitivity of LBNE 
improves more for LO (lower octant) than HO (higher octant), for any assumed true hierarchy. In Fig. 1 below we plot CP asymmetry,
\begin{equation}
A_{CP} = \frac{P(\nu_{\mu}\rightarrow \nu_{e})-P(\overline{\nu_{\mu}}\rightarrow \overline{\nu_{e}})}{P( \nu_{\mu}\rightarrow \nu_{e})+P(\overline{\nu_{\mu}}\rightarrow \overline{\nu_{e}})}
\label{diseqn}
\end{equation}
as a function of leptonic CPV phase $ \delta_{CP} $, for 0 $\leq \delta_{CP} \leq 2 \pi$.
It was shown in \cite{DK} that, using near detector (and combining with reactor experiments) at LBNE, the sentivity to measure CPV
phase (and hence CP asymmetry) improves more at lower octant of $ \theta_{23} $. CP asymmetry also depends on the mass hierarchy.
For NH, CP asymmetry is more in LO than in HO. For IH, CP asymmetry is more in LO than in HO. In this work we have used above information to calculate dependance of leptogenesis on octant of $ \theta_{23}$ and quadrant of CPV phase.
\begin{figure}[b]
\centerline{\includegraphics[width=7.8cm]{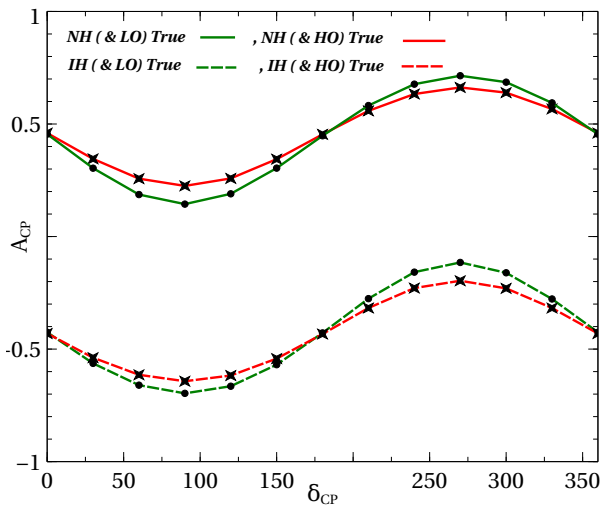}}
\caption{CP asymmetry vs $\delta_{CP}$ at DUNE/LBNE, for both the hierarchies. In Fig. 1 red and green solid (dotted) lines are for NH (IH) with types of curve to distinguish HO and LO as the true octant respectively.\label{pcv.pdf}}
\end{figure}
From Fig. 1 we see that
\par  
\begin{equation}
A_{CP}(LO) > A_{CP}(HO)
\label{diseqn}
\end{equation}
\par
For a given true hierarchy, there are eight degenerate solutions    
$$\delta_{CP}(\text{first quadrant}) - \theta_{23}(\text{lower octant})$$
$$\delta_{CP}(\text{second quadrant}) - \theta_{23}(\text{lower octant})$$
$$\delta_{CP}(\text{third quadrant}) - \theta_{23}(\text{lower octant})$$
$$\delta_{CP}(\text{fourth quadrant}) - \theta_{23}(\text{lower octant})$$
$$\delta_{CP}(\text{first quadrant}) - \theta_{23}(\text{higher octant})$$
$$\delta_{CP}(\text{second quadrant}) - \theta_{23}(\text{higher octant})$$
$$\delta_{CP}(\text{third quadrant}) - \theta_{23}(\text{higher octant})$$
\begin{equation}
\delta_{CP}(\text{fourth quadrant}) - \theta_{23}(\text{higher octant})
\label{diseqn}
\end{equation}
This eight-fold degeneracy can be viewed as 
\begin{equation}
\text{Quadrant of CPV phase} - \text{Octant of}\hspace{.1cm}  \theta_{23}
\label{diseqn}
\end{equation}
entanglement. Out of these eight degenerate solutions, only one should be true solution. To pinpoint one true solution, this entanglement has to be broken. We have shown \cite{DK} that sensitivity to discovery potential of CPV at LBNEs in LO is improved more, if data from near detector of LBNEs, or from Reactor experiments is added to data from FD of LBNEs as shown in Fig. 3 of \cite{DK}. Therefore 8-fold degeneracy of (3) gets reduced to 4-fold degeneracy, with our proposal \cite{DK}. Hence, following this 4-fold degeneracy still remains to be resolved.
 \begin{equation*}
\delta_{CP}(\text{first quadrant}) - \theta_{23}(\text{LO})
\end{equation*}
\begin{equation*}
\delta_{CP}(\text{second quadrant}) - \theta_{23}(\text{LO})
\end{equation*}
\begin{equation*}
\delta_{CP}(\text{third quadrant}) - \theta_{23}(\text{LO})
\end{equation*}
\begin{equation}
\delta_{CP}(\text{fourth quadrant}) - \theta_{23}(\text{LO})
\end{equation}

The possibility of $ \theta_{23} > 45^{0}$, ie HO of $ \theta_{23}$ is also considered in this work. In this context the degeneracy is
\begin{equation*}
\delta_{CP}(\text{first quadrant}) - \theta_{23}(\text{HO})
\end{equation*}
\begin{equation*}
\delta_{CP}(\text{second quadrant}) - \theta_{23}(\text{HO})
\end{equation*}
\begin{equation*}
\delta_{CP}(\text{third quadrant}) - \theta_{23}(\text{HO})
\end{equation*}
\begin{equation}
\delta_{CP}(\text{fourth quadrant}) - \theta_{23}(\text{HO})
\end{equation}
 
\begin{figure}[b]
\centerline{\begin{subfigure}[]{\includegraphics[width=6.8cm]{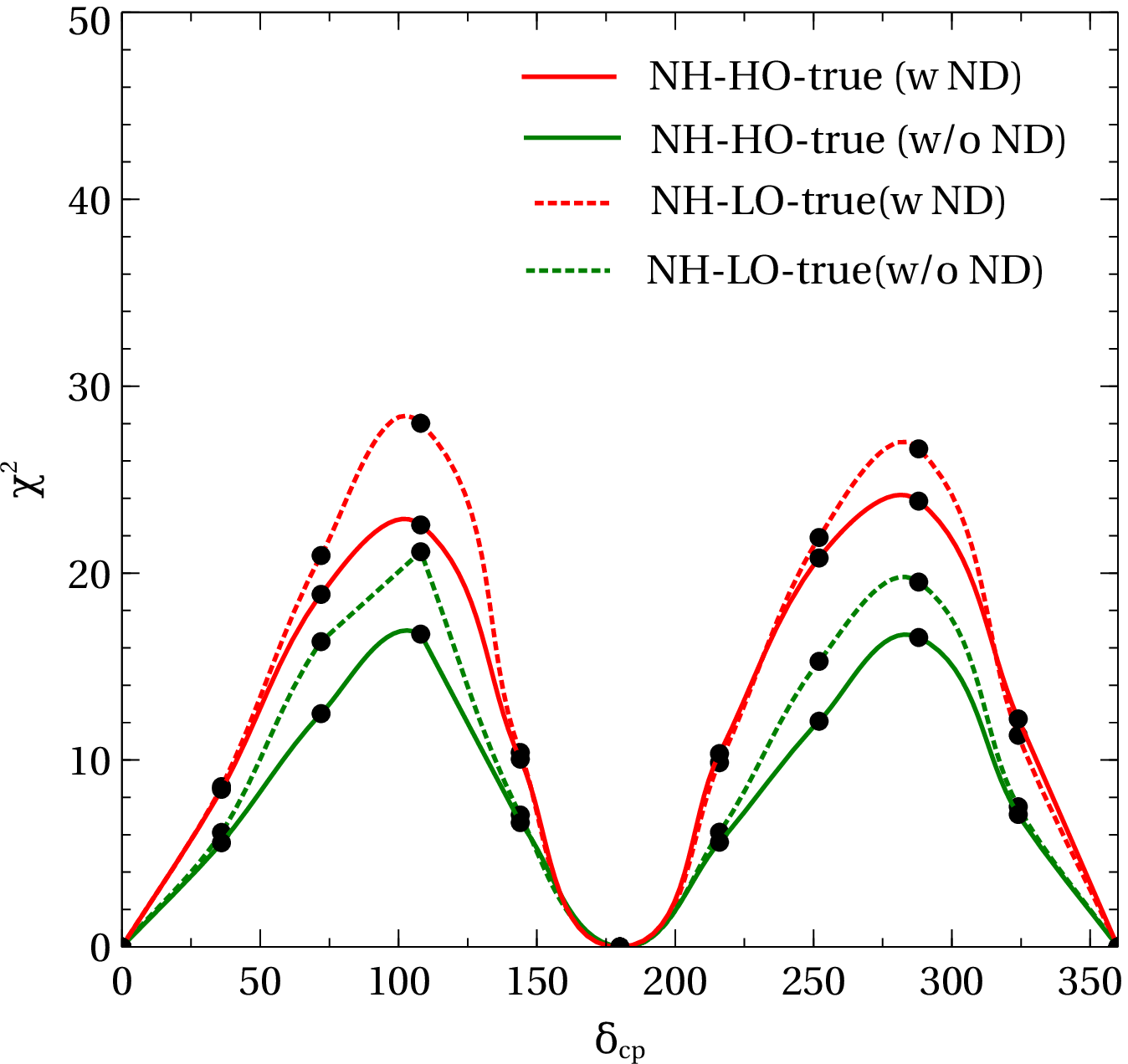}}\end{subfigure}
\begin{subfigure}[]{\includegraphics[width=6.8cm]{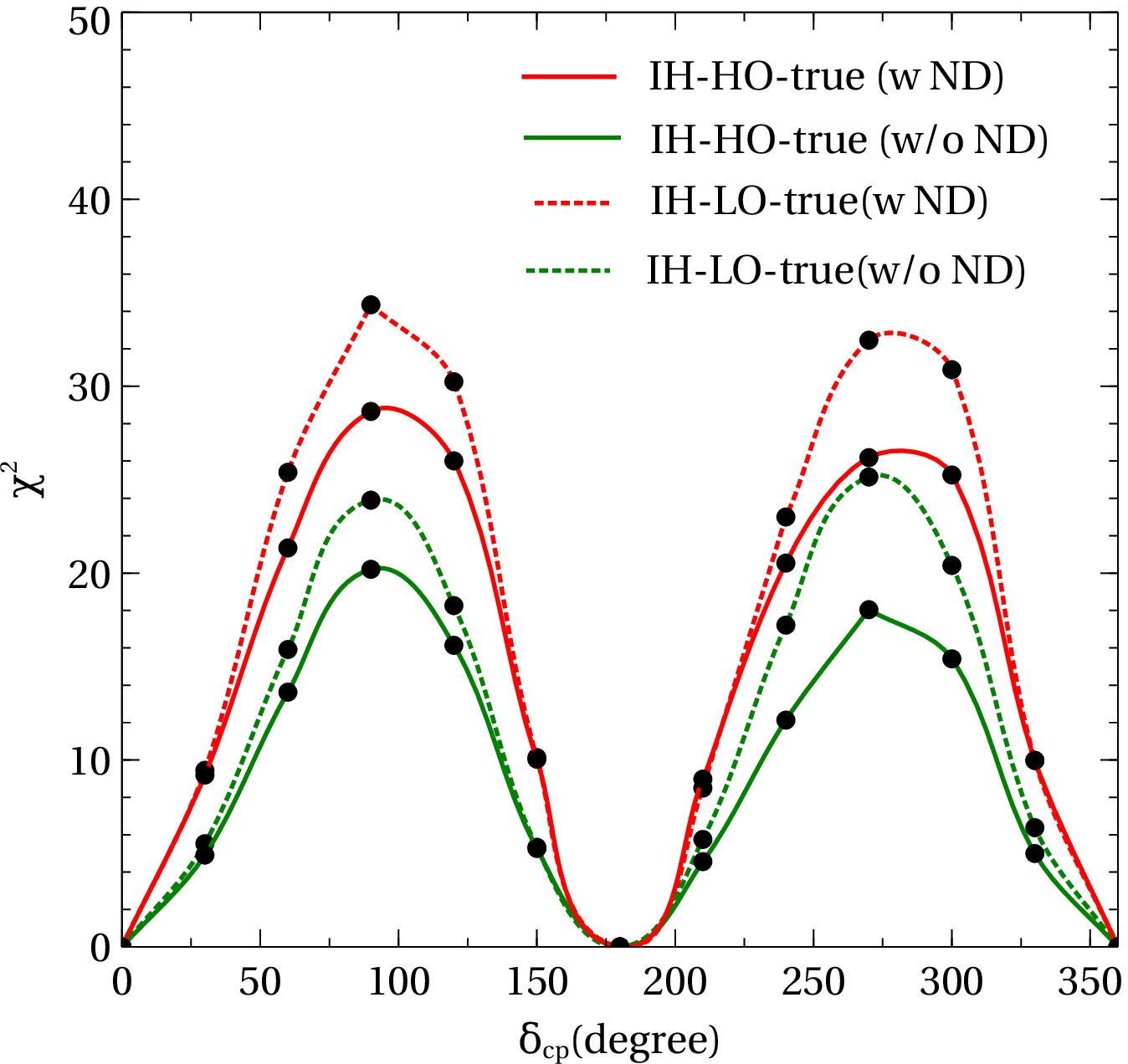}}\end{subfigure}}
\caption{In Fig. 2a and 2b $\delta_{CP}$ Vs $\Delta \chi^{2}$ sensitivity corresponding to CP discovery potential at LBNEs, for both the hierarchies and octant is shown.}
\end{figure}

In this work, we propose that leptogenesis can be used to break above mentioned 4-fold degeneracy of Eq. (5),(6). It is known that observed baryon asymmetry of the Universe (BAU) can be explained via leptogenesis \cite{Bari, RM, Bhupal, Pd, GC}. In leptogenesis, the lepton-antilepton asymmetry can be explained, if there are complex yukawa couplings or complex fermion mass matrices. This in turn arises due to complex leptonic CPV phases, $\delta_{CP}$, in fermion mass matrices. If all other parameters except leptonic $\delta_{CP}$ phase in the formula for lepton - antilepton asymmetry are fixed, for example, then observed value of BAU from experimental observation can be used to constrain quadrant of $\delta_{CP}$, and hence 4-fold entanglement of (5),(6) can be broken. An experimental signature of CP violation associated to the dirac phase $\delta_{CP}$, in PMNS matrix \cite{Mki}, can in principle be obtained, by searching for CP asymmetry in $ \nu $ flavor oscillation. To elucidate this proposal, we consider model independent scenario, in which BAU arises due to leptogenesis, and this lepton-antilepton asymmetry \cite{Uma} is generated by the out of equilibrium decay of the right handed, heavy majorana neutrinos, which form an integral part of seesaw mechanism for neutrino masses and mixings. Since our proposal is model independent, we consider type I seesaw mechanism, just for simplicity.

\section{\textbf{Leptogenesis and Baryogenesis in Type I Seesaw SO(10) Models}}

In Grand Unified theories like SO(10), one right handed heavy majorana neutrino per generation is added to Standard Model \cite{l,m,n,ibarra}, and they couple to left handed $ \nu $ via Dirac mass matrix $m_{D}$. When the neutrino mass matrix is diagonalised, we get two eigen values $ - $ light neutrino $ \sim $ $ \frac{m^{2}_{D}}{M_{R}} $ and a heavy neutrino state $ \sim $ $M_{R}$. This is called type I See Saw mechanism. Here, decay of the lightest of the three heavy RH majorana neutrinos, $M_{1}$, i.e $M_{3}, M_{2}\gg M_{1}$ will contribute to $l-\bar{l}$ asymmetry (for leptogenesis), i.e $\epsilon^{CP}_{l}$. In the basis where RH $\nu$ mass matrix is diagonal, the type I contribution to $\epsilon^{CP}_{l}$ is given by decay of $M_{1}$
 \begin{equation}
\epsilon^{CP}_{l}=\frac{\Gamma(M_{1}\rightarrow lH) - \Gamma(M_{1}\rightarrow \bar{l} \bar{H})}{\Gamma(M_{1}\rightarrow lH) + (\Gamma(M_{1}\rightarrow \bar{l} \bar{H})}, 
\end{equation}
where $\Gamma(M_{1}\rightarrow lH)$ means decay rate of heavy majorana RH $\nu$ of mass $M_{1}$ to a lepton and Higgs. We assume a 
normal mass hierarchy for heavy Majorana neutrinos. In this scenario the lightest of heavy Majorana neutrinos is in thermal equilibrium
 while the heavier neutrinos, $M_{2}$ and $ M_{3} $, decay. Any asymmetry produced by the out of equilibrium decay of $M_{2}$ and 
$ M_{3} $ will be washed away by the lepton number violating interactions mediated by $ M_{1} $. Therefore, the final lepton-antilepton asymmetry is given only by the CP-violating decay of $ M_{1} $ to standard model leptons (l) and Higgs (H). This contribution 
is \cite{Mh}:
\begin{equation}
\epsilon_{l}=-\frac{3M_{1}}{8\pi}\frac{Im[\Delta m^{2}_{\odot}R^{2}_{12}+\Delta m^{2}_{A}R^{2}_{13}]}{\upsilon^{2}\sum 
|R_{ij}|^{2}m_{j}}. 
\end{equation}
$R$ is a complex orthogonal matrix with the property that $RR^{T} = 1$. $R$ can be parameterized as \cite{Osc}:
\begin{equation}
R = D_{\sqrt{M^{-1}}}Y_{\nu}UD_{\sqrt{K^{-1}}},
\end{equation}
where $Y_{\nu}$ is the matrix of neutrino yukawa couplings. In the basis, where the charged-lepton Yukawa matrix, $Y_{e}$ and gauge interactions are flavour-diagonal,
$ D_{K} = U^{T}KU  $, where $K=Y_{\nu}^{T}M_{R}^{-1}Y_{\nu}$. $U$ is the PMNS matrix and  $M_{R}$ is the RH neutrino Majorana scale. In the basis of right handed neutrinos,  $D_{M} = Diag(M_{1}, M_{2},M_{3})$ where $M_{3}, M_{2}\gg M_{1}$. Equation (8) relates the lepton asymmetry to both the solar ($\Delta m^{2}_{21}$) and atmospheric ($ \Delta m^{2}_{A} $) mass squared differences. 
Thus the magnitude of the matter-antimatter asymmetry can be predicted in terms of low energy oscillation parameters, $\Delta m^{2}_{21}$, $ \Delta m^{2}_{A} $ and a CPV phase. Here matrix $R$ is dependent on both $U_{PMNS}$ and $V_{CKM}$, and it can be shown that,
\begin{eqnarray*}
\text{Im}R^{2}_{13}& = &-\text{Sin}(2\delta_{q})\text{Cos}^{2}(\theta^{l}_{23})\text{Cos}^{2}(\theta^{l}_{13})\text{Sin}^{2}(\theta^{q}_{13})-2\text{Sin}(\delta_{q})\text{Cos}(\theta_{13}^{q})\text{Cos}(\theta_{23}^{l})\text{Cos}^{2}(\theta^{l}_{13})\text{Sin}(\theta_{12}^{q})\text{Sin}(\theta_{13}^{q})\text{Sin}(\theta_{23}^{l})\\& & +2\text{Sin}(-\delta_{l}-\delta_{q})\text{Cos}(\theta_{12}^{q})\text{Cos}(\theta_{13}^{q})\text{Cos}(\theta^{l}_{23})\text{Cos}(\theta_{13}^{l})\text{Sin}(\theta_{13}^{q})\text{Sin}(\theta_{13}^{l})-2\text{Sin}(\delta_{l})\text{Cos}(\theta_{12}^{q})\text{Cos}^{2}(\theta_{13}^{q})\text{Cos}(\theta^{l}_{13})\\& &\text{Sin}(\theta_{12}^{q})\text{Sin}(\theta_{23}^{l})\text{Sin}(\theta_{13}^{l})-\text{Sin}(2\delta_{l})\text{Cos}^{2}(\theta_{12}^{q})\text{Cos}^{2}(\theta_{13}^{q})\text{Sin}^{2}(\theta^{l}_{13})-2\text{Sin}(\delta_{l})\text{Cos}^{2}(\theta_{12}^{q})\text{Cos}^{2}(\theta_{13}^{q})\text{Sin}^{2}(\theta^{l}_{13})
\end{eqnarray*}
\begin{eqnarray*}
\text{Im}R^{2}_{12}&=&2\text{Sin}(\delta_{q})\text{Cos}(\theta_{13}^{q})\text{Cos}^{2}(\theta_{12}^{l})\text{Cos}(\theta^{l}_{23})\text{Sin}(\theta_{12}^{q})\text{Sin}(\theta_{13}^{q})\text{Sin}(\theta_{23}^{l})+2\text{Sin}(\delta_{q})\text{Cos}(\theta_{12}^{q})\text{Cos}(\theta_{13}^{q})\text{Cos}(\theta^{l}_{12})\text{Cos}(\theta^{l}_{13})\\& & \text{Sin}(\theta_{13}^{q})\text{Sin}(\theta_{12}^{q})\text{Sin}(\theta_{23}^{l})-\text{Sin}(2\delta_{q})\text{Cos}^{2}(\theta_{12}^{l})\text{Sin}^{2}(\theta_{13}^{q})\text{Sin}^{2}(\theta^{l}_{23})-2\text{Sin}(\delta_{l}-\delta_{q})\text{Cos}(\theta_{13}^{q})\text{Cos}(\theta_{12}^{l})\text{Cos}^{2}(\theta^{l}_{23})\\& &\text{Sin}(\theta^{q}_{12})\text{Sin}(\theta_{13}^{q})\text{Sin}(\theta_{12}^{l})\text{Sin}(\theta_{13}^{l})-2\text{Sin}(\delta_{l}-\delta_{q})\text{Cos}(\theta_{12}^{q})\text{Cos}(\theta_{13}^{q})\text{Cos}(\theta^{l}_{23})\text{Cos}(\theta^{l}_{13})\text{Sin}(\theta_{13}^{q})\text{Sin}^{2}(\theta_{12}^{l})\text{Sin}(\theta_{23}^{l})\\& &-2\text{Sin}(\delta_{l})\text{Cos}^{2}(\theta_{13}^{q})\text{Cos}(\theta_{12}^{l})\text{Cos}(\theta^{l}_{23})\text{Sin}^{2}(\theta^{q}_{12})\text{Sin}(\theta_{12}^{l})\text{Sin}(\theta_{23}^{l})\text{Sin}(\theta_{13}^{l})+2\text{Sin}(\delta_{l}-2\delta_{q})\text{Cos}(\theta_{12}^{l})\text{Cos}(\theta_{23}^{l})\\& &\text{Sin}^{2}(\theta^{q}_{13})\text{Sin}(\theta^{l}_{12})\text{Sin}(\theta_{23}^{l})\text{Sin}(\theta_{13}^{l})-2\text{Sin}(\delta_{l})\text{Cos}(\theta_{12}^{q})\text{Cos}^{2}(\theta_{13}^{q})\text{Cos}(\theta^{l}_{13})\text{Sin}(\theta^{q}_{12})\text{Sin}^{2}(\theta_{12}^{l})\text{Sin}(\theta_{23}^{l})\text{Sin}(\theta_{13}^{l})\\& &+ 2\text{Sin}(\delta_{l}-\delta_{q})\text{Cos}(\theta_{13}^{q})\text{Cos}(\theta_{12}^{l})\text{Sin}(\theta^{q}_{12})\text{Sin}(\theta^{q}_{12})\text{Sin}(\theta_{12}^{l})\text{Sin}^{2}(\theta_{23}^{l})\text{Sin}(\theta^{l}_{13})+  2\text{Sin}(2\delta_{l}-2\delta_{q})\text{Cos}^{2}(\theta_{23}^{l})\\ & &\text{Sin}^{2}(\theta_{13}^{q})\text{Sin}^{2}(\theta^{l}_{12})\text{Sin}^{2}(\theta^{l}_{13})+  2\text{Sin}(2\delta_{l}-\delta_{q})\text{Cos}(\theta_{13}^{q})\text{Cos}(\theta_{23}^{l})\text{Sin}(\theta^{q}_{12})\text{Sin}(\theta^{q}_{13})\text{Sin}^{2}(\theta_{12}^{l})\text{Sin}(\theta_{23}^{l})\text{Sin}^{2}(\theta^{l}_{13})\\& & + \text{Sin}(2\delta_{l})\text{Cos}^{2}(\theta_{13}^{q})\text{Sin}^{2}(\theta_{12}^{q})\text{Sin}^{2}(\theta^{l}_{12})\text{Sin}^{2}(\theta^{l}_{23})\text{Sin}^{2}(\theta_{13}^{l})
\end{eqnarray*}
\begin{eqnarray*}
R_{11}&=&\text{Cos}(\theta_{12}^{q})\text{Cos}(\theta_{13}^{q})\text{Cos}(\theta_{12}^{l})\text{Cos}(\theta_{13}^{l})+e^{-i\delta_{q}}\text{Sin}(\theta_{13}^{q})\text{Sin}(\theta_{12}^{l})\text{Sin}(\theta_{23}^{l})-e^{-i\delta_{l}}e^{-i\delta_{q}}\text{Sin}(\theta_{13}^{q})\text{Cos}(\theta_{12}^{l})\text{Cos}(\theta_{23}^{l})\text{Sin}(\theta_{13}^{l})\\ & & -\text{Cos}(\theta_{13}^{q})\text{Sin}(\theta_{12}^{q})\text{Cos}(\theta_{23}^{l})\text{Sin}(\theta_{12}^{l}) - e^{-i\delta_{l}}\text{Cos}(\theta_{13}^{q})\text{Sin}(\theta_{12}^{q})\text{Cos}(\theta_{12}^{l})\text{Sin}(\theta_{23}^{l})\text{Sin}(\theta_{13}^{l})
\end{eqnarray*}
\begin{eqnarray*}
R_{12}&=&\text{Cos}(\theta_{12}^{q})\text{Cos}(\theta_{13}^{q})\text{Cos}(\theta_{13}^{l})\text{Sin}(\theta_{12}^{l})-e^{-i\delta_{q}}\text{Sin}(\theta_{13}^{q})\text{Cos}(\theta_{12}^{l})\text{Sin}(\theta_{23}^{l})-e^{-i\delta_{l}}e^{-i\delta_{q}}\text{Cos}(\theta_{23}^{l})\text{Sin}(\theta_{12}^{l})\text{sin}(\theta_{13}^{l})\text{Sin}(\theta_{13}^{}q)\\ & & -\text{Cos}(\theta_{13}^{q})\text{Sin}(\theta_{12}^{q})\text{Cos}(\theta_{12}^{l})\text{Cos}(\theta_{23}^{l}) - e^{-i\delta_{l}}\text{Cos}(\theta_{13}^{q})\text{Sin}(\theta_{12}^{q})\text{Sin}(\theta_{12}^{l})\text{Sin}(\theta_{23}^{l})\text{Sin}(\theta_{13}^{l})
\end{eqnarray*}
\begin{equation}
R_{13} = e^{-i\delta_{q}}\text{Cos}(\theta_{23}^{l})\text{Cos}(\theta_{13}^{l})\text{Sin}(\theta_{13}^{q})-\text{Cos}(\theta_{13}^{q})\text{Cos}(\theta_{13}^{l})\text{Sin}(\theta_{12}^{q})\text{Sin}(\theta_{23}^{l}) - e^{-i\delta_{l}}\text{Cos}(\theta_{12}^{q})\text{cos}(\theta_{13}^{q})\text{Sin}(\theta_{13}^{l})
\end{equation}
where, $\theta^{l}_{23}$, $\theta^{l}_{13}$, $\theta^{l}_{12}$ denote the three $\nu$ mixing angles,  $\theta^{q}_{23}$,
 $\theta^{q}_{13}$, $\theta^{q}_{12}$ are the quark mixing angles. $\delta_{l}$ and $\delta_{q}$ are the leptonic CPV phase and quark
 CPV phase respectively. When left-right symmetry is broken at high intermediate mass scale $ M_{R} $ in SO(10) theory,
 CP asymmetry is given by
\begin{equation}
\epsilon_{l}=-\frac{3M_{1}}{8\pi}\frac{Im[\Delta m^{2}_{A}R^{2}_{13}]}{\upsilon^{2}\sum 
|R_{ij}|^{2}m_{j}} 
\end{equation}
where
$$|R_{11}|^{2}=\text{Cos}^{2}(\theta_{12}^{l})\text{Cos}^{2}(\theta_{13}^{l}), |R_{12}|^{2} = \text{Sin}^{2}(\theta_{12}^{l})\text{Cos}^{2}(\theta_{13}^{l}), |R_{13}|^{2} = \text{Cos}^{2}(\delta_{l})\text{Sin}^{2}(\theta_{13}^{l})+ \text{Sin}^{2}(\delta_{l})\text{Sin}^{2}(\theta_{13}^{l})$$and
\begin{equation}
\text{Im}R^{2}_{13} = -\text{Sin}^{2}(2\delta_{l})\text{Sin}^{2}(\theta_{13}^{l})
\end{equation}
The neutrino oscillation data used in our numerical calculations are summarised as follows {\cite{Frero}}. 
$$\Delta m^{2}_{21}[10^{-5}eV^{2}] = 7.62 \pm 0.19$$ 
$$|\Delta m^{2}_{31}|[10^{-3}eV^{2}] = 2.55^{+0.06}_{-0.09}(2.43^{+0.07}_{-0.06})$$ 
$$\text{Sin}^{2}\theta_{12} = 0.320^{+0.016}_{-0.017} $$
$$\text{Sin}^{2}\theta_{23} = 0.613^{+0.022}_{-0.040}(0.600^{+0.026}_{-0.031})$$ 
\begin{equation}
\text{Sin}^{2}\theta_{13} = 0.0246^{+0.0049}_{-0.0028}(0.0250^{+0.0026}_{-0.0027})
\end{equation}
For $\Delta m^{2}_{31}, Sin^{2}\theta_{23}, Sin^{2}\theta_{13}$, the quantities inside the bracket corresponds to inverted neutrino mass hierarchy and those outside the bracket corresponds to normal mass hierarchy. The errors are within the 1$\sigma$ range of the $\nu$ oscillation parameters. It may be noted that some results on neutrino masses and mixings using updated values of running quark and lepton masses in
SUSY SO(10) have also been presented in \cite{Gayatri}. Though we consider 3-flavour neutrino scenario, 4-flavour neutrinos with 
sterile neutrinos as fourth flavour, are also possible \cite{KBo}. It is worth mentionng that $ \nu $ masses and mixings can lead 
to charged lepton flavor violation in grand unified theories like SO(10) \cite{GG}.
\par
The origin of the baryon asymmetry in the universe (baryogenesis) is a very interesting topic of current research. A well known mechanism is the baryogenesis via leptogenesis, where the out-of-equilibrium decays of heavy right-handed Majorana neutrinos produce a lepton asymmetry which is transformed into a baryon asymmetry by electroweak sphaleron processes \cite{Hooft, Manton, Kuzmin}. Lepton asymmetry is partially converted to baryon asymmetry through B+L violating sphaleron interactions \cite{ME}. As proposed in \cite{SO}, a baryon 
asymmetry can be generated from a lepton asymmetry. The baryon asymmetry is defined as:
\begin{equation}
Y_{B} = \frac{n_{B}-n_{\bar{B}}}{s}= \frac{n_{B}-n_{\bar{B}}}{7n_{\gamma}}=\frac{\eta_{B}}{7}, 
\end{equation}
where $n_{B}, n_{\bar{B}},n_{\gamma}$ are number densities of baryons, antibaryons and photons respectively, $s$ is the entropy density,  $ \eta $ is the baryon-to-photon ratio, $ 5.7 \times 10^{-10} \leq \eta_{B} \leq 6.7 \times 10^{-10}$ (95 \%  C.L) \cite{B.D}. The lepton number is converted into the baryon number through electroweak sphaleron process \cite{Hooft,Manton,Kuzmin}.

\begin{equation}
Y_{B} = \frac{a}{a-1}Y_{L}, a = \frac{8N_{F} + 4N_{H}}{22N_{F}+13N_{H}},
\end{equation}
where $ N_{f} $ is the number of families and $ N_{H} $ is the
number of light Higgs doublets. In case of SM, $N_{f} = 3$ and $ N_{H} = 1 $. The lepton asymmetry is as follows:
\begin{equation}
Y_{L} = d\frac{\epsilon_{l}}{g^{*}}.
\end{equation}
$d$ is a dilution factor and $g^{*} = 106.75$ in the standard case \cite{SO}, is the effective number of degrees of freedom. The dilution factor d \cite{SO} is, $d = \frac{0.24}{k(lnk)^{0.6}}$ for $k\geq 10 $ and $d = \frac{1}{2k}, d = 1$ for $1\leq k \leq 10$ and $0\leq k \leq 1$ respectively, where the parameter k \cite{SO} is, $ k=\frac{M_{P}}{1.7\upsilon^{2}32\pi\sqrt{g^{*}}}\frac{(M_{D}\dagger M_{D})_{11}}{M_{1}} $, here $ M_{P} $ is the Planck mass. We have used the form of Dirac neutrino mass matrix $M_{D}$ from \cite{Joshipura}.

\section{\textbf{Analysis And Discussion Of Results}}
For our numerical analysis, we take the current experimental data for three neutrino
mixing angles as inputs, which are given at $1\sigma$ $-$ $3\sigma$ C.L, as presented in \cite{Frero}. Here, we perform numerical analysis and present
results both for normal hierarchy, inverted hierarchy, HO, LO from Fig. 2.
We have explored the CP asymmetry using Eq. (7)-Eq. (12) and corresponding baryon asymmetry using Eq. (14)-(16), for 152 different combinations (shown in Table I-XII) of the two hierarchies (NH and IH), two types of octants$-$ LO and HO, w ND, w/o ND (with and without near detector) and $\delta_{CP}$ corresponding to maximum $\chi^{2}$ (for maximum sensitivity from Fig. 2(a), 2(b)), for which the CP discovery potential of the DUNE is maximum. We also consider non maximal values of $ \delta_{CP} $ corresponding to $  \chi^{2} = $ 4, 9, 16, 25 from Fig. 2. We examine these different cases in the light of recent ratio of the baryon to photon density bounds, $ 5.7 \times 10^{-10} \leq \eta_{B} \leq 6.7 \times 10^{-10}$ (CMB), and checked for which of the 152 cases, our calculated value of $|\eta_{B}|$ lies within this range. 
 
\begin{table}[H]
\begin{center}
\begin{tabular}{|c|c|c|c|c|c|}

\hline 
\textbf{Case}& \textbf{hierarchy, octant} & \textbf{w ND/\hspace{.1cm}OR w/o ND} & \textbf{$\delta_{CP}$}  & \textbf{$\epsilon_{l}$}& \textbf{$|\eta_{B}|$}\\ 
\hline 
$1$ &$\text{NH, LO}$ & $WND$&$ 101 $ & $-.0000177532$&$7.39703\times10^{-8} $   \\
\hline 
$2$&$\text{NH, LO}$ & $WND$&$ 280$ & $-.0000125002$&$5.17312\times10^{-8}$ \\
\hline
$3$&$\text{NH, LO}$ & $W/oND$&$ 108 $ & $.0000153489$&$6.35202\times10^{-8} $\\
 \hline
$4$&$\text{NH, LO}$ & $W/oND$&$ 282 $ & $7.53352\times10^{-6}$&$3.11769\times10^{-8} $\\
\hline
$5$&$\text{IH, LO}$ & $W/ND$&$ 83 $ & $2.56383\times10^{-6}$&$1.06102\times10^{-8} $ \\
\hline
$6$&$\text{IH, LO}$ & $W/ND$&$ 276 $ & $1.01403\times10^{-6}$&$4.19647\times10^{-9} $ \\
\hline
$7$&$\text{IH, LO}$ & $W/oND$&$ 88 $ & $1.46427 \times 10^{-7}$&$6.05975\times10^{-10} $\\
 \hline
$8$&$\text{IH, LO}$ & $W/oND$&$ 275 $ & $3.6845\times10^{-6}$&$1.5248\times10^{-8} $\\
\hline
\end{tabular}
\end{center}
\caption{Calculated values of CP asymmetry $\epsilon_{l}$ and baryon to photon ratio $|\eta_{B}|$ in case of LO, for $ R_{1j} $ elements of R matrix consisting of $ U_{PMNS} $ and $V_{CKM}$ for the values of $ \delta_{CP} $ when the CP discovery potential of the LBNE/DUNE is maximum as shown in Fig. 2.}

\end{table} 
\begin{table}[H]
\begin{center}
\begin{tabular}{|c|c|c|c|c|c|}

\hline 
\textbf{Case}&\textbf{hierarchy, Octant} & \textbf{w ND/ OR w/o ND} & \textbf{$\delta_{CP}$}  & \textbf{$\epsilon_{l}$}& \textbf{$|\eta_{B}|$}\\ 
\hline 
$1$&$\text{NH, LO}$ & $WND$&$ 101 $ & $.0000268767$&$1.11227\times10^{-7} $   \\
\hline 
$2$&$\text{NH, LO}$ & $WND$&$ 280$ & $.0000238272$&$9.86068\times10^{-8} $ \\
\hline
$3$&$\text{NH, LO}$ & $W/oND$&$ 108 $ & $.0000231986$&$9.60055\times10^{-8} $\\
 \hline
$4$&$\text{NH, LO}$ & $W/oND$&$ 282 $ & $.0000332106$&$1.3744\times10^{-7} $\\
\hline
$5$&$\text{IH, LO}$ & $WND$&$ 83 $ & $.0000109298$&$4.5232\times10^{-8} $ \\
\hline
$6$&$\text{IH, LO}$ & $W/ND$&$ 276 $ & $-3.3319\times10^{-6}$&$1.37888\times10^{-8} $ \\
\hline
$7$&$\text{IH, LO}$ & $W/oND$&$ 88 $ & $2.96234\times10^{-7}$&$1.22594\times10^{-9} $\\
 \hline
$8$&$\text{IH, LO}$ & $W/oND$&$ 270 $ & $-9.18963\times10^{-7}$&$3.80305\times10^{-9} $\\
\hline
\end{tabular}
\end{center}
\caption{Same as in Table I, except here R matrix consists of $ U_{PMNS} $ only. }
\end{table}
 
We find that out of 32 different  cases corresponding to maximal sensitivity $ \chi^{2} $ (from Fig. 2) as shown in Table I$-$IV, our calculated value of BAU is larger than the currently allowed range of BAU  except for two cases: case 7 in table I and case 5 in table III for which the calculated $|\eta_{B}|$ is compatible with the present range of baryon to photon density ratio \cite{B.D}. In Table I, case 7 which has $\delta_{CP}= 88^{0}$ or $0.488\pi $\hspace{.1cm}(first quadrant), IH and atmospheric angle $\theta_{23}$ in LO has $ \eta_{B}  = 6.05975 \times  10^{-10} $, consistent with 
its best fit value, $ \eta_{B}  = 6.05 \times  10^{-10} $ \cite{B.D}. For this case, $\epsilon_{l} = 1.46427\times 10^{-7}$  lies within the Davidson and Ibbara bounds \cite{ibarra}, ($\epsilon_{l}\leq 4.59 \times 10^{-5} $). In Table III, case 5 has $\delta_{CP}= 95^{0}$ or $0.5277 \pi$ \hspace{.1cm}(second quadrant), IH and atmospheric angle $\theta_{23}$ in HO has BAU equal to  $6.2157 \times  10^{-10} $ which is in accord with the present $|\eta_{B}|$ bounds and it leads to CP asymmetry $|\epsilon_{l}| = 1.50195\times 10^{-7}$ that lies within the Davidson and Ibarra bounds.

\begin{figure}[b]
\centerline{\begin{subfigure}[]{\includegraphics[width=9.7cm]{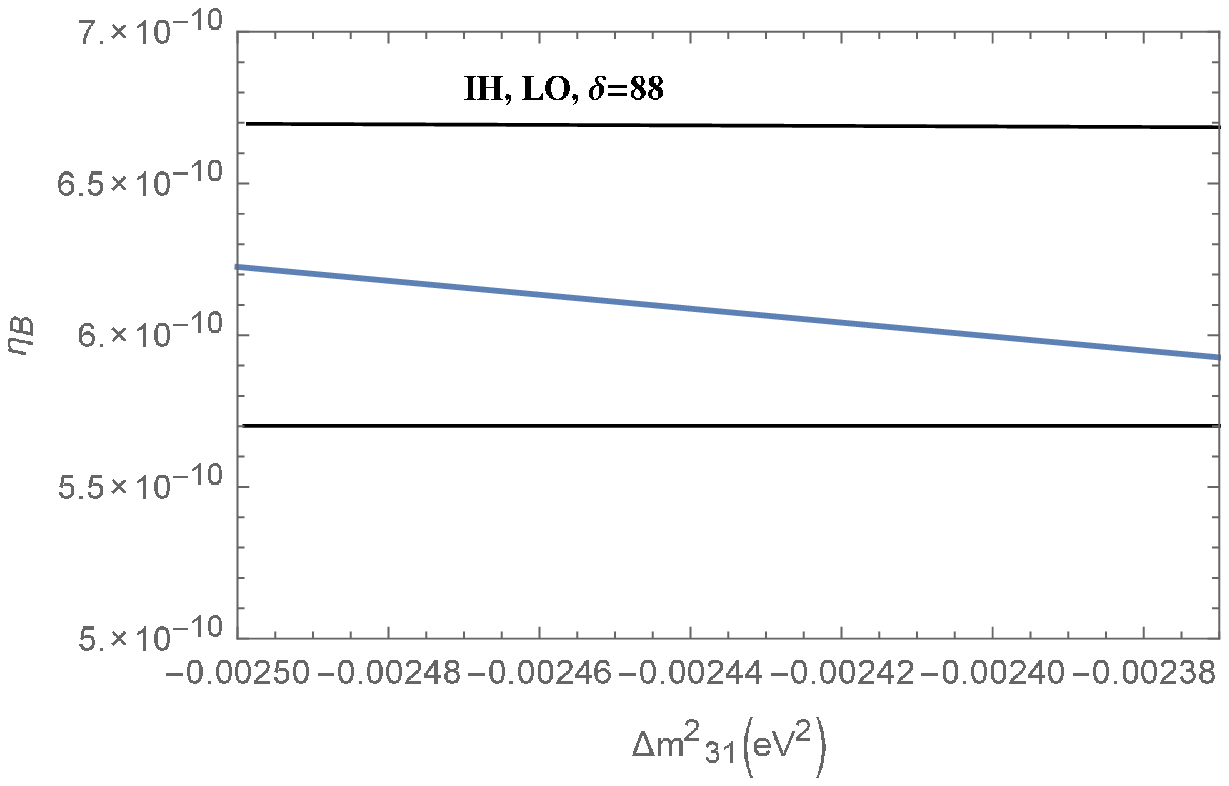}}\end{subfigure}
\begin{subfigure}[]{\includegraphics[width=9.7cm]{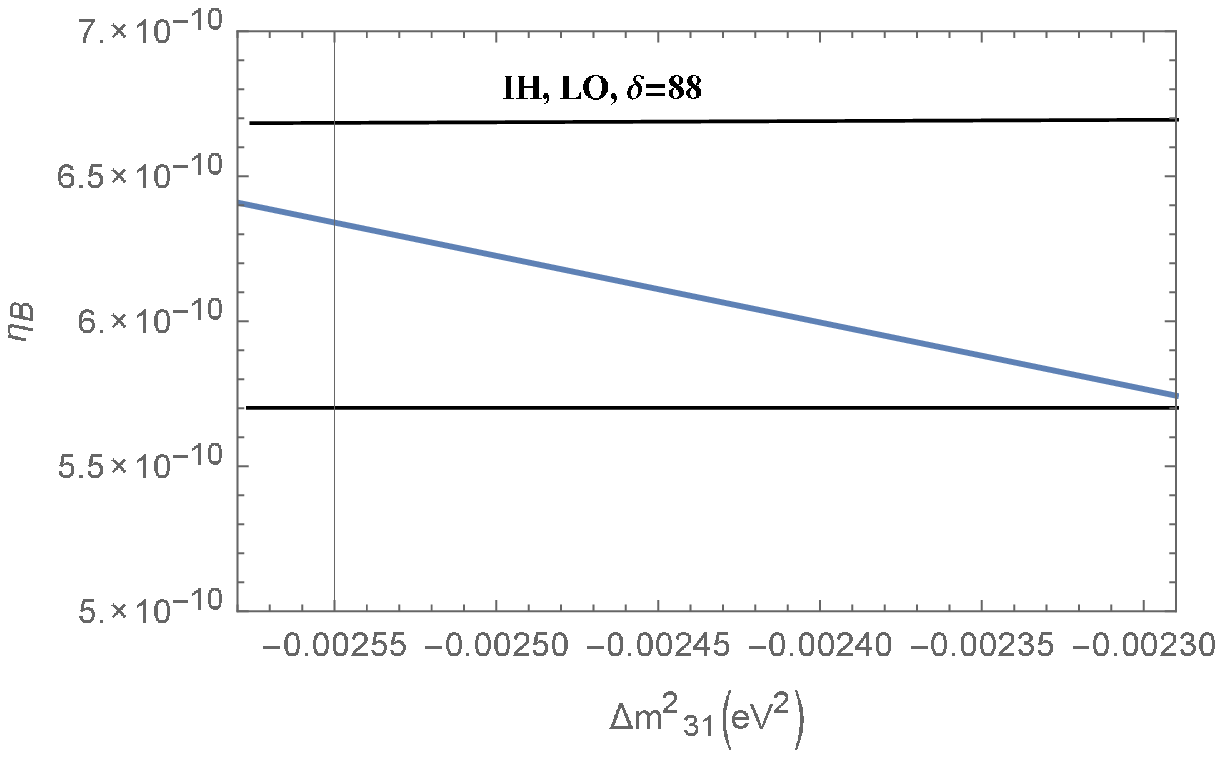}}\end{subfigure}}
\caption{Variation of $ \eta_{B} $ with $ \Delta m^{2}_{31} $, for case 7 of Table I based on 1$ \sigma $ and 2$\sigma $ range of $ \Delta m^{2}_{31} $ in Fig. 3(a) and 3(b) respectively. Plot of $ \eta_{B} $ Vs $ \Delta m^{2}_{31} $[$eV^{2}$] with CP phases $ \delta_{CP}= 0.488 \pi$ for the case when R matrix consists of both $V_{CKM}$ and $U_{PMNS}$. The blue solid line in Fig. 3(a), 3(b) corresponds to $\theta_{23}$ in LO, 
 $ \delta_{CP}= 0.488 \pi $\hspace{.1cm}(first quadrant) and IH. The black horizontal line corresponds to the upper and lower limit on $ \eta_{B} $, $5.7\times 10^{-10} < \eta_{B} < 6.7 \times 10^{-10} $. As can be seen from the figure, the plots in Fig. 3(a), 3(b) satisfy the current experimental constraints on $ \eta_{B} $.}
\end{figure}

Figure 3 shows the allowed regions of $ |\eta_{B}| $ in the plane charted by ($ \Delta m^{2}_{31} $, $ |\eta_{B}| $) for $ \delta_{CP} $ allowed at maximal sensitivity of CP discovery potential from Fig. 2 (case 7 of Table I). Here we show the variation of $ |\eta_{B}| $ with $ \Delta m^{2}_{31} $, taking the variation of the later within its 1$ \sigma  $ and 2$ \sigma $ limits. It can be seen that $ \eta_{B} $ for our calculation 
(blue solid line) lies within the result of its global fit value ($5.7\times 10^{-10} < \eta_{B} < 6.7 \times 10^{-10} $) shown in \cite{B.D}.

\begin{table}[H]
\begin{center}
\begin{tabular}{|c|c|c|c|c|c|}

\hline 
\textbf{Case}& \textbf{hierarchy, octant} & \textbf{w ND/\hspace{.1cm}w/o ND} & \textbf{$\delta_{CP}$}  & \textbf{$\epsilon_{l}$}& \textbf{$|\eta_{B}|$}\\ 
\hline 
$1$ &$\text{NH, HO}$ & $WND$&$ 101 $ & $-.0000189857$ & $7.85709\times10^{-8} $   \\
\hline 
$2$&$\text{NH, HO}$ & $WND$&$ 281$ & $-3.51289\times10^{-5}$&$1.45378\times10^{-7}$ \\
\hline
$3$&$\text{NH, HO}$ & $W/oND$&$ 102 $ & $2.72017 \times 10^{-5}$& $1.12572 \times 10^{-7} $\\
 \hline
$4$&$\text{NH, HO}$ & $W/oND$&$ 283 $ & $-1.82461\times10^{-5}$&$7.551\times10^{-8} $\\
\hline
$5$&$\text{IH, HO}$ & $W/ND$&$ 95 $ & $-1.50195\times10^{-7}$&$6.2157\times10^{-10} $ \\
\hline
$6$&$\text{IH, HO}$ & $W/ oND$&$ 94 $ & $-8.7785\times10^{-8}$&$3.63291\times10^{-10} $ \\
\hline
$7$&$\text{IH, HO}$ & $W/ND$&$ 281 $ & $-5.8547 \times 10^{-6}$&$2.42292\times10^{-8} $\\
 \hline
 $8$&$\text{IH, HO}$ & $W/oND$&$ 272$ & $9.97129\times10^{-6}$&$4.12654\times10^{-8} $\\
\hline
\end{tabular}
\end{center}
\caption{Same as in Table I, but HO values are used.}
\end{table}

\begin{table}[H]
\begin{center}
\begin{tabular}{|c|c|c|c|c|c|}

\hline 
\textbf{Case}& \textbf{hierarchy, octant} & \textbf{w ND/\hspace{.1cm}w/o ND} & \textbf{$\delta_{CP}$}  & \textbf{$\epsilon_{l}$}& \textbf{$|\eta_{B}|$}\\ 
\hline 
$1$ &$\text{NH, HO}$ & $WND$&$ 101 $ & $.0000268767$ & $1.11227\times10^{-7} $   \\
\hline 
$2$&$\text{NH, HO}$ & $WND$&$ 281$ & $.0000112743$&$4.66576\times10^{-8}$ \\
\hline
$3$&$\text{NH, HO}$ & $W/oND$&$ 102 $ & $6.73637\times 10^{-6}$& $2.78779 \times 10^{-8} $\\
 \hline
$4$&$\text{NH, HO}$ & $W/oND$&$ 283 $ & $.0000163668$&$6.77325\times10^{-8} $\\
\hline
$5$&$\text{IH, HO}$ & $W/ND$&$ 95 $ & $4.1771\times10^{-6}$&$1.72891\times10^{-8} $ \\
\hline
$6$&$\text{IH, HO}$ & $W/ oND$&$ 94 $ & $-1.99098\times10^{-6}$&$8.23952\times10^{-9} $ \\
\hline
$7$&$\text{IH, HO}$ & $W/ND$&$ 281 $ & $7.65022 \times 10^{-6}$&$3.16598\times10^{-8} $\\
 \hline
 $8$&$\text{IH, HO}$ & $W/oND$&$ 272$ & $-.00001093$&$4.52369\times10^{-8} $\\
\hline
\end{tabular}
\end{center}
\caption{Same as in Table III, but R matrix consists of $ U_{PMNS} $ only. }
\end{table}

Next, we explore values of $ \delta_{CP} $ corresponding to $\chi^{2} $ = 4, 9, 16, 25 from Fig. 2 for which the CP discovery potential of the LBNE/DUNE is non maximal. For $\chi^{2} $ = 2 $ \sigma $, 3$ \sigma $ sensitivity of the CP discovery potential, Table V-VIII summarise the results where we find that out of the 64 possible cases in all, for 63 cases the calculated BAU is larger than the currently allowed range of BAU \cite{B.D} by almost two to three orders of magnitude except for case 4 of Table VII where $ \delta _{CP} $ = 1.924 $ \pi $,  IH, HO, has BAU of the order of 8.65034 $ \times 10^{-12}$  less than the allowed $|\eta_{B}| $ limit.
\par 
We examine 56 possible cases for non maximal CP discovery sensitivity potential of the LBNE/DUNE from Fig. 2 summarised in Table IX-XII corresponding to $ \chi^{2}$ at 4$\sigma$, 5$\sigma$ C.L out of which only 3 cases are consistent with the experimental results of $ |\eta_{B}| $ bounds, (a) Case 15 of Table XI where $ \frac{\delta_{CP}}{\pi}=1.43 $, $ \nu $ mass spectrum of IH nature, atmospheric angle $ \theta_{23} $ in HO, has CP asymmetry $ \epsilon _{l} = 1.48671 \times 10^{-7} $ which lies within $ |\epsilon _{l}^{max}| = 4.59 \times 10^{-5} $ (Davidson Ibbara bounds) and $ |\eta_{B}| = 6.15262 \times 10^{-10}$ that agrees with the present BAU range. It is worth noting that this value of $ \frac{\delta_{CP}}{\pi}=1.43 $ is close to the central value of $ \delta_{CP} $ from the recent global fit result \cite{kol}, (b) Case 13 of Table XI that locates $ \frac{\delta_{CP}}{\pi}= 0.3833 $, $ \nu $ mass spectrum of IH nature, $ \theta_{23} $ in HO, $ \epsilon _{l} = 1.40342 \times 10^{-7} $ ($ \leq $ $ |\epsilon _{l}^{max}| = 4.59 \times 10^{-5} $) has $ |\eta_{B}| = 5.80973 \times 10^{-10}$, consistent with the allowed BAU range. Here  $ R_{1j} $ elements of R matrix consists of $ U_{PMNS} $ and $V_{CKM}$ in both the cases above, (c) Case 4 of Table XII which has $ \frac{\delta_{CP}}{\pi}= 1.727 $, IH  $ \nu $ mass spectrum, $ \theta_{23} $ in LO, $ |\epsilon _{l}| = 1.47958 \times 10^{-7} $ lies within $ |\epsilon _{l}^{max}| = 4.59 \times 10^{-5} $ and $ |\eta_{B}| = 6.12311 \times 10^{-10}$ that agrees with the current experimental constraints \cite{B.D}. 
\begin{table}[H]
\begin{center}
\begin{tabular}{|c|c|c|c|c|c|}

\hline 
\textbf{Case}&\textbf{hierarchy, Octant} & \textbf{$\Delta \chi^{2}$} & \textbf{$\delta_{CP}$}  & \textbf{$\epsilon_{l}$}& \textbf{$|\eta_{B}|$}\\ 
\hline 
$1$&$\text{NH, LO}$ & $4$&$ 17 $ & $-3.57328\times10^{-5}$&$1.4787\times10^{-7} $   \\
\hline 
$2$&$\text{NH, LO}$ & $9$&$ 38 $ & $1.91122\times10^{-5}$&$7.90943\times10^{-8} $ \\
\hline
$3$&$\text{NH, LO}$ & $9$&$ 147 $ & $3.19392\times10^{-5}$&$1.32178\times10^{-7} $ \\
\hline
$4$&$\text{NH, LO}$ & $4$&$ 154$ & $3.33001\times10^{-6}$&$1.3781\times10^{-8} $ \\
\hline
$5$&$\text{NH, LO}$ & $4$&$ 203.5 $ & $3.31724\times10^{-5}$&$1.37281\times10^{-7} $\\
\hline
$6$&$\text{NH, LO}$ & $9$&$ 213 $ & $1.18422\times10^{-5}$&$4.9008\times10^{-8} $\\
 \hline
$7$&$\text{NH, LO}$ & $9$&$ 332$ & $-7.01565\times10^{-6}$&$2.90337\times10^{-8} $\\
\hline
$8$&$\text{NH, LO}$ & $4$&$ 346.5$ & $1.72854\times10^{-6}$&$7.15341\times10^{-9} $\\
\hline
$9$&$\text{NH, HO}$ & $4$&$ 17 $ & $-3.6128\times10^{-5}$&$1.49513\times10^{-7} $   \\
\hline 
$10$&$\text{NH, HO}$ & $4$&$ 155 $ & $-2.65416\times10^{-5}$&$1.0984\times10^{-7} $ \\
\hline
$11$&$\text{NH, HO}$ & $4$&$ 203 $ & $3.76207\times10^{-5}$&$1.5569\times10^{-7} $ \\
\hline
$12$&$\text{NH, HO}$ & $4$&$ 347.5$ & $8.3309\times10^{-7}$&$3.44768\times10^{-9} $ \\
\hline
$13$&$\text{NH, HO}$ & $9$&$ 38.3 $ & $-1.54969\times10^{-5}$&$6.41328\times10^{-8} $\\
\hline
$14$&$\text{NH, HO}$ & $9$&$ 147 $ & $3.09483\times10^{-5}$&$1.28077\times10^{-7} $\\
 \hline
$15$&$\text{NH, HO}$ & $9$&$ 212$ & $-3.30145\times10^{-5}$&$1.36628\times10^{-7} $\\
\hline
$16$&$\text{NH, HO}$ & $9$&$ 333.5$ & $-1.97211\times10^{-6}$&$8.16144\times10^{-9} $\\
\hline
\end{tabular}
\end{center}
\caption{Calculated values of CP asymmetry $\epsilon_{l}$ and baryon to photon ratio $|\eta_{B}|$ in case of NH, for $ R_{1j} $ elements of R matrix consisting of $ U_{PMNS} $ and $V_{CKM}$ for DUNE/LBNE with its near detector, with $ \chi^{2} = 4 \hspace{.1cm}\text{and}\hspace{.1cm}9 $  measuring CP discovery sensitivity from Fig. 2.}
\end{table}

\begin{center}
\begin{figure*}[htbp]
\centering
{\begin{subfigure}[]{\includegraphics[width=8.87 cm]{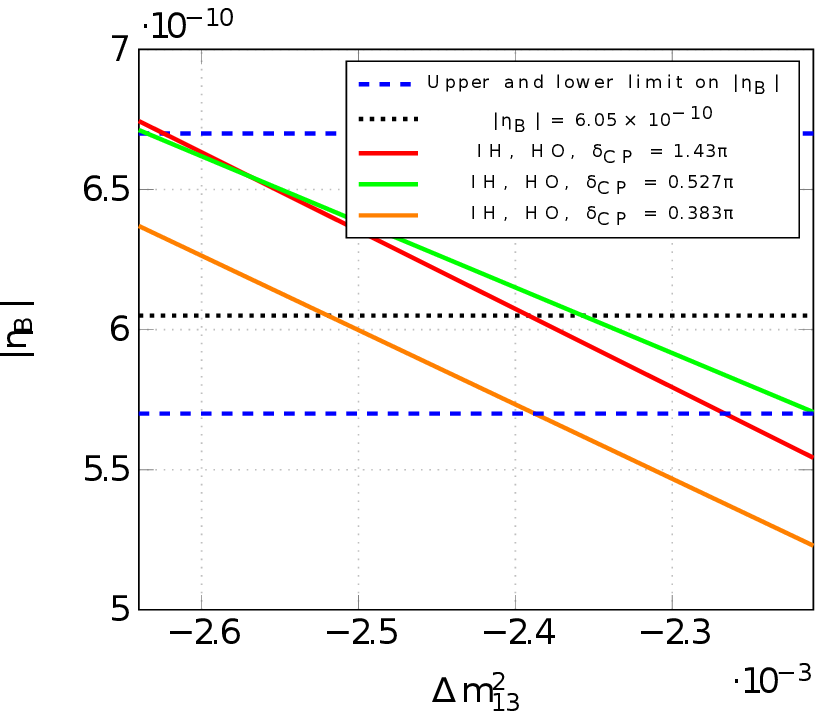}}\end{subfigure}
\begin{subfigure}[]{\includegraphics[width=8.87cm]{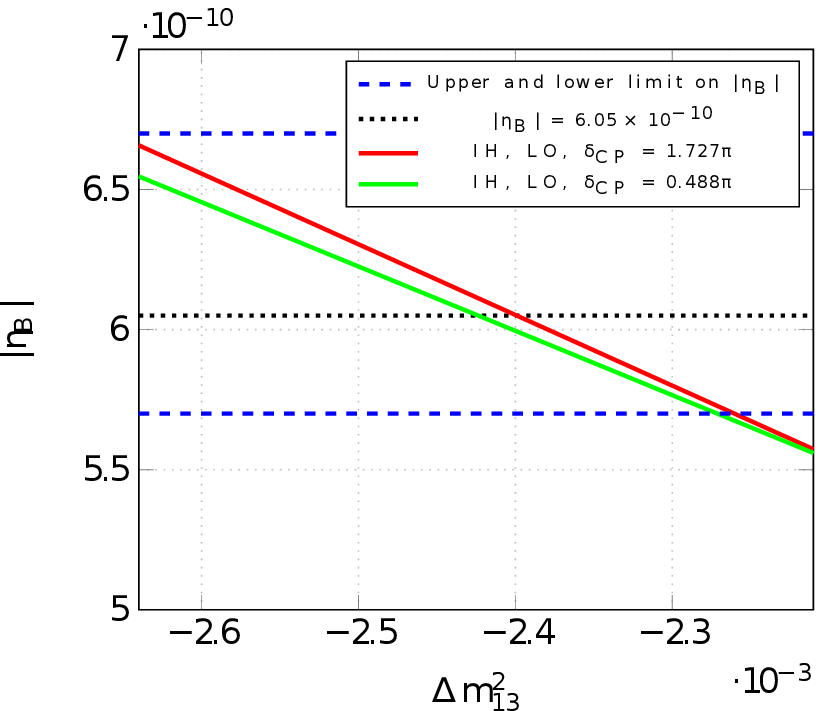}}\end{subfigure}\\
\caption{Variation of $ \eta_{B} $ with $ \Delta m^{2}_{31} $ within its 3$ \sigma $ C.L.  The upper and lower limit on $ \eta_{B} $, $5.7\times 10^{-10} < \eta_{B} < 6.7 \times 10^{-10} $ are characterised by blue dashed horizontal lines. Black dotted line corresponds to  best fit value, $ \eta_{B} = 6.05 \times 10^{-10}$. In the left panel, Fig. 4(a) shows the plot of $ \eta_{B} $ Vs $ \Delta m^{2}_{31} $ for $ \delta_{CP} = 1.43 \pi, 0.527 \pi, 0.383 \pi $. Fig. 4(b) of  right panel frames the variation of $ \eta_{B} $ with $ \Delta m^{2}_{31} $ for $ \delta_{CP} = 0.488 \pi, 1.727 \pi$.}}
\end{figure*}
\end{center}

Plugging the experimental data for $ \Delta m^{2} _{31} $ at 3$ \sigma $  C.L, and other $ \nu $ oscillation parameters at best fit into Eq. (8 - 12) we predict the values of $ \eta_{B} $ from Eq. (14, 15, 16) as shown in the Fig. 4. The figure displays the allowed regions of $| \eta_{B} | $ in the plane ($ \Delta m^{2}_{13}, | \eta_{B} | $) for experimental results of $ \Delta m^{2}_{31} $ at 3$ \sigma $ C.L. In Fig. 4(a) red solid line conforms to the case 15 of Table XI, where $ \delta_{CP} = 1.43 \pi $, $ \nu $ mass spectrum of IH structure, atmospherc angle $ \theta_{23} $ in HO and $ |\eta_{B}| $ in the range consistent with  $5.7\times 10^{-10} < \eta_{B} < 6.7 \times 10^{-10} $ except for $ \Delta m^{2}_{31} >- 2.2695 \times 10^{-3} eV^{2} $ and  $ \Delta m^{2}_{31} <- 2.635 \times 10^{-3} eV^{2} $ where the red solid line departs from the experimental bound on $ \eta_{B} $. The orange solid line in Fig. 4(a) depicts case 13 of Table XI which has $ \delta_{CP} = 0.383 \pi $, $ \nu $ mass structure of IH spectrum, $ \theta_{23} $ in HO and $ |\eta_{B}| $ in the allowed range followed by the experimental constraints on $ |\eta_{B}| $ except for $ \Delta m^{2}_{31} >- 2.385 \times 10^{-3} eV^{2} $. Slight variation of $ \eta_{B} $ for  $ \delta_{CP} = 0.5277 \pi $ can be seen from Fig. 4(a) for $ \Delta m^{2}_{31} < -2.63\times 10^{-3} eV^{2}$ (green solid line). Similarly the green solid line in Fig. 4(b) corresponds to $ \delta_{CP} =0.488 \pi $, IH $ \nu $ spectrum, which is consistent with the allowed range of BAU for $ \Delta m^{2}_{31} < -2.27\times 10^{-3} eV^{2}$. The red solid line in Fig. 4(b) characterises case 4 of Table XII, which has $ \delta_{CP} = 1.727 \pi $, $ \nu $ mass structure of IH nature, atmospherc angle $ \theta_{23} $ in LO and $ |\eta_{B}| $ in the range favoured by the present experimental limit on $ |\eta_{B}| $, $5.7\times 10^{-10} < |\eta_{B}| < 6.7 \times 10^{-10} $ except for $ \Delta m^{2}_{31} >- 2.255 \times 10^{-3} eV^{2} $ where the curve fails to fall in the allowed $ |\eta_{B}| $ bounds even at $ 2 \sigma $ C.L of $ \Delta m^{2}_{31} $.
\par 
From the above discussion, we conclude that, out of total 152 cases presented in Table I-XII, only for five cases, the values of $ \eta_{B} $ lie within the experimental limits, which are summarised in Table XIII.
\begin{table}[H]
\begin{center}
\begin{tabular}{|c|c|c|c|c|c|}

\hline 
\textbf{Case}&\textbf{hierarchy, Octant} & \textbf{$\Delta \chi^{2}$} & \textbf{$\delta_{CP}$}  & \textbf{$\epsilon_{l}$}& \textbf{$|\eta_{B}|$}\\ 
\hline 
$1$&$\text{NH, LO}$ & $4$&$ 17 $ & $1.76335\times10^{-5}$&$7.2975\times10^{-8} $   \\
\hline 
$2$&$\text{NH, LO}$ & $9$&$ 38 $ & $1.88675\times10^{-5}$&$7.80817\times10^{-8} $ \\
\hline
$3$&$\text{NH, LO}$ & $9$&$ 147 $ & $-3.2199\times10^{-5}$&$1.33253\times10^{-7} $ \\
\hline
$4$&$\text{NH, LO}$ & $4$&$ 203.5$ & $-3.28826\times10^{-6}$&$1.36082\times10^{-7} $ \\
\hline
$5$&$\text{NH, LO}$ & $4$&$ 154 $ & $4.1195\times10^{-6}$&$1.70482\times10^{-8} $\\
\hline
$6$&$\text{NH, LO}$ & $9$&$ 213 $ & $-3.16969\times10^{-5}$&$1.31175\times10^{-7} $\\
 \hline
$7$&$\text{NH, LO}$ & $9$&$ 332$ & $-3.00567\times10^{-5}$&$1.24385\times10^{-7} $\\
\hline
$8$&$\text{NH, LO}$ & $4$&$ 346.5$ & $3.20414\times10^{-5}$&$1.32601\times10^{-7} $\\
\hline
$9$&$\text{NH, HO}$ & $4$&$ 17 $ & $1.76335\times10^{-5}$&$7.2975\times10^{-8} $   \\
\hline 
$10$&$\text{NH, HO}$ & $4$&$ 155 $ & $2.83588\times10^{-5}$&$1.1736\times10^{-7} $ \\
\hline
$11$&$\text{NH, HO}$ & $4$&$ 203 $ & $3.10849\times10^{-5}$&$1.28642\times10^{-7} $ \\
\hline
$12$&$\text{NH, HO}$ & $4$&$ 347.5$ & $-2.16746\times10^{-5}$&$8.96988\times10^{-8} $ \\
\hline
$13$&$\text{NH, HO}$ & $9$&$ 38.3 $ & $3.10849\times10^{-5}$&$1.28642\times10^{-7} $\\
\hline
$14$&$\text{NH, HO}$ & $9$&$ 147 $ & $-3.2199\times10^{-5}$&$1.33253\times10^{-7} $\\
 \hline
$15$&$\text{NH, HO}$ & $9$&$ 212$ & $3.82461\times10^{-6}$&$1.58278\times10^{-8} $\\
\hline
$16$&$\text{NH, HO}$ & $9$&$ 333.5$ & $2.77229\times10^{-7}$&$1.14729\times10^{-7} $\\
\hline
\end{tabular}
\end{center}
\caption{Same as in Table V, but $ R = U_{PMNS} $ only.}
\end{table}
 
\begin{center}
\begin{figure*}[htbp]
\centering
{\begin{subfigure}[]{\includegraphics[width=8.7cm]{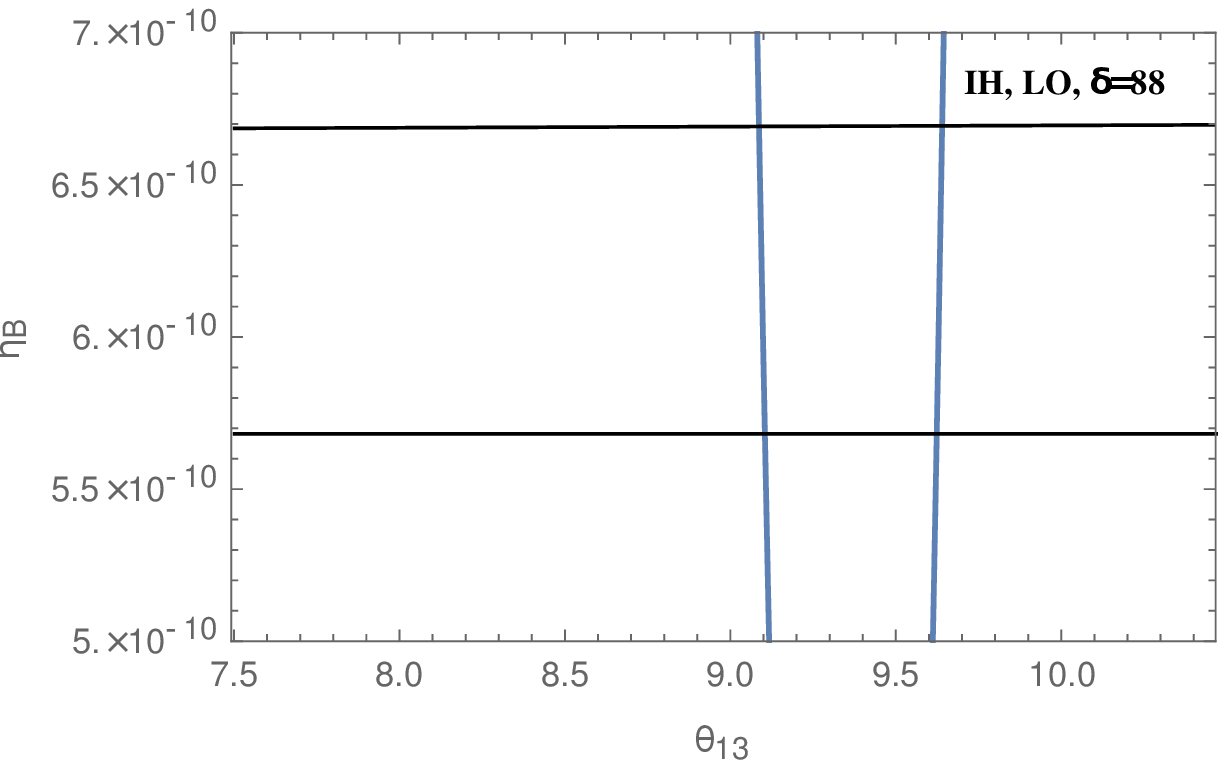}}\end{subfigure}
\begin{subfigure}[]{\includegraphics[width=8.7cm]{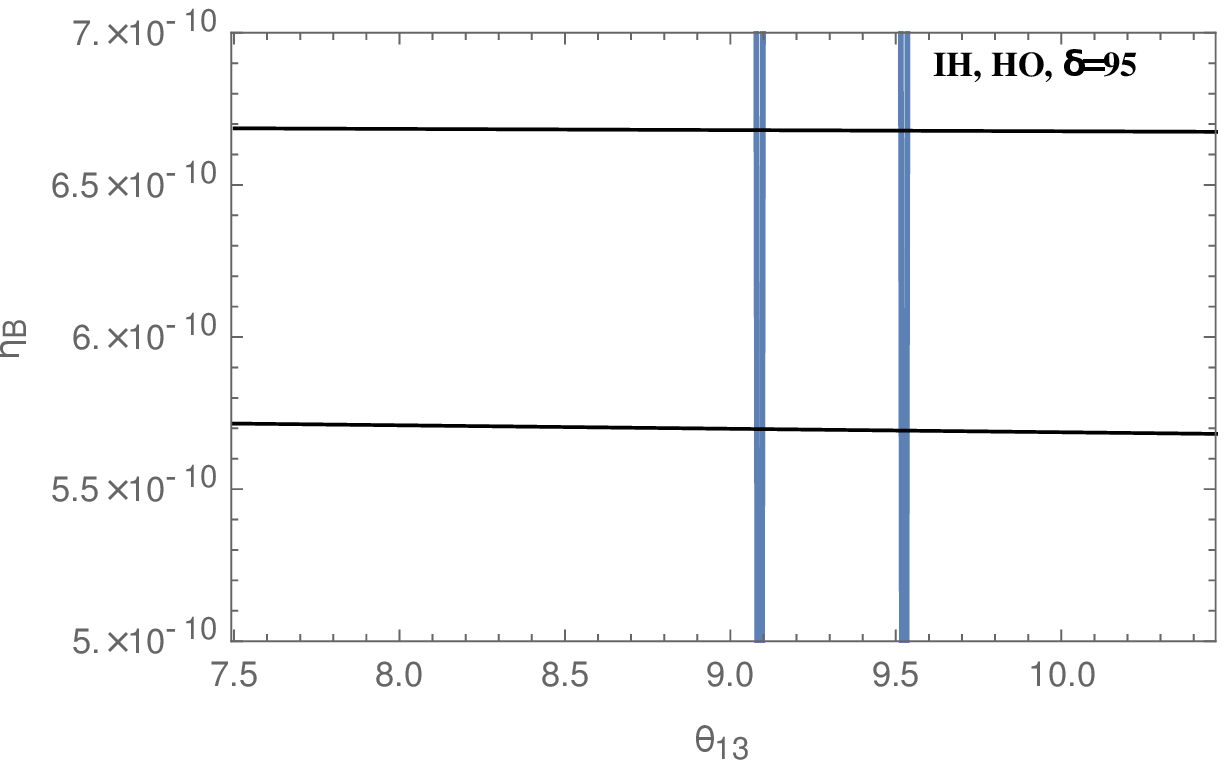}}\end{subfigure}\\
\begin{subfigure}[]{\includegraphics[width=8.7cm]{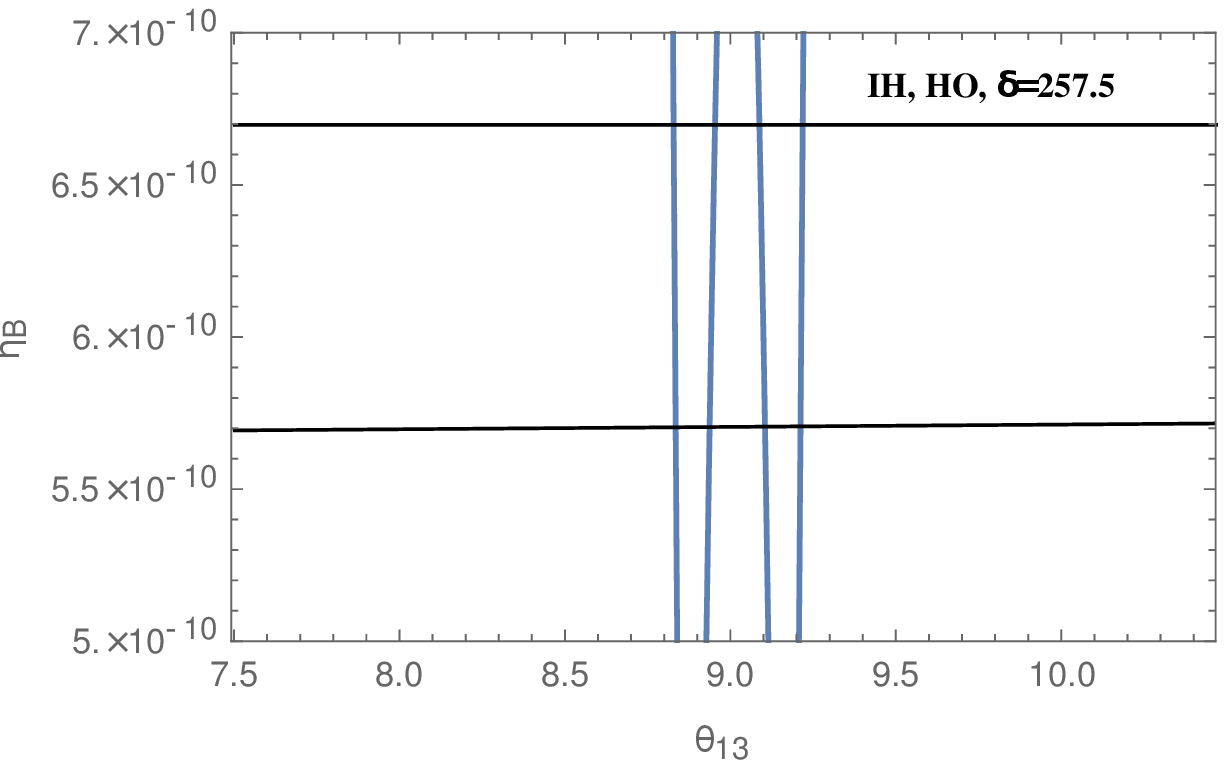}}\end{subfigure}
\begin{subfigure}[]{\includegraphics[width=8.7cm]{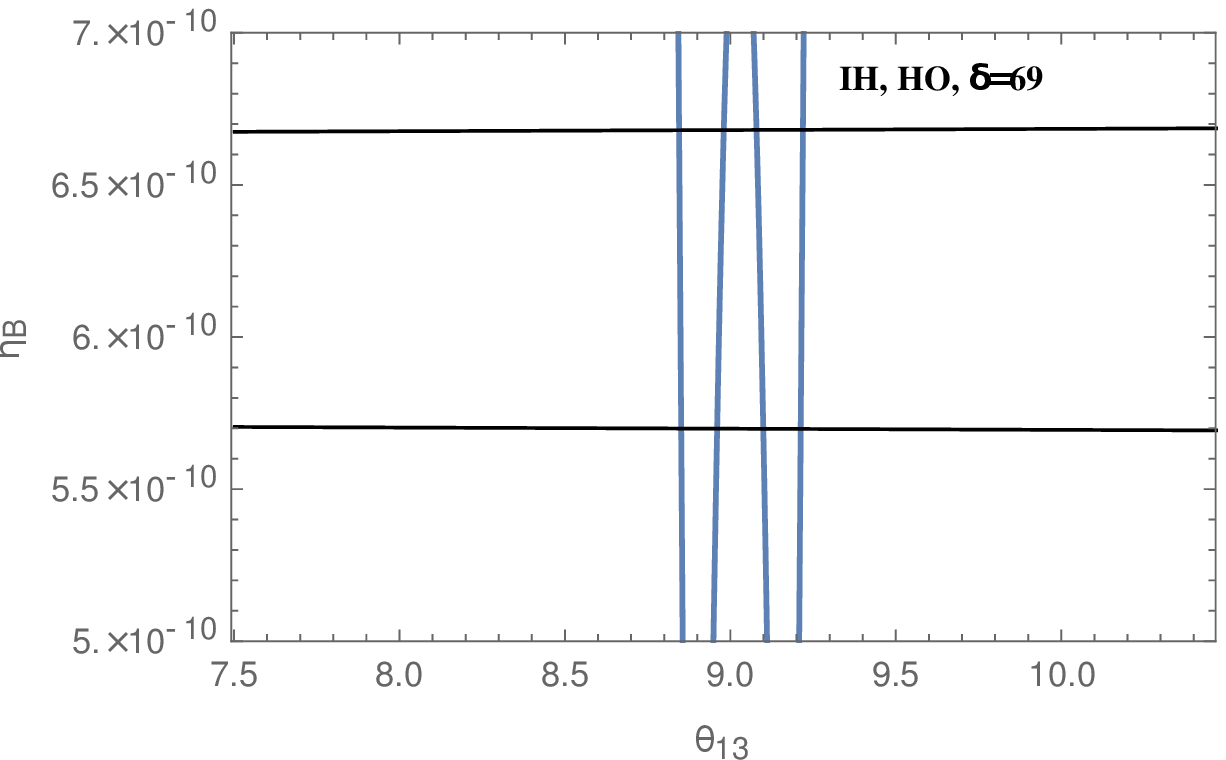}}\end{subfigure}\\
\begin{subfigure}[]{\includegraphics[width=8.7cm]{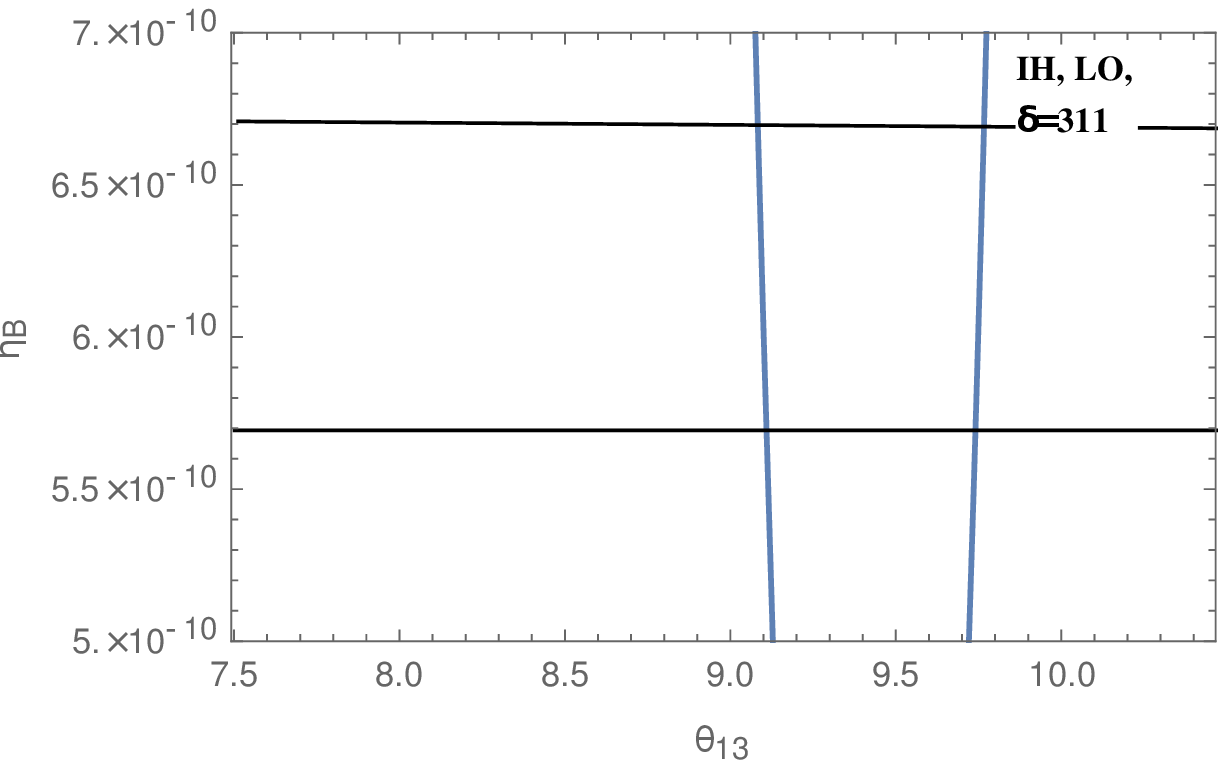}}\end{subfigure}\\
\caption{Plot of $ \eta_{B} $ vrs $ \theta_{13} $ with CP phases in Fig. 5(a) $ \delta_{CP}= 88^{0} $, IH, LO; in Fig. 5(b) $ \delta_{CP}= 95 ^{0}$, IH, HO; in Fig. 5(c) $ \delta_{CP}= 257.5^{0}$, IH, HO; in Fig. 5(d) $ \delta_{CP}= 69 ^{0}$, IH, HO and in Fig. 5(e) $ \delta_{CP}= 311 ^{0}$, IH, LO within the 3 $\sigma$ errors of the best fit values of $ \theta_{13} $ for the favoured cases. The black solid  horizontal line corresponds to the upper and lower limit on $ \eta_{B} $, $5.7\times 10^{-10} < \eta_{B} < 6.7 \times 10^{-10} $. }}
\end{figure*}
\end{center}

Figure 5 completes our discussion by showing the allowed regions in the plane ($ \theta_{13}, |\eta_{B}| $) which is done for five cases favoured by our analysis above. The shapes of the curves are somewhat symmetrical in Fig. 5(c) and 5(d) about $ \theta_{13} = 9^{0}$ for $ \delta_{CP} = 1.43 \pi$, IH, $ \theta_{23} $ in HO and  $ \delta_{CP} = 0.383 \pi$, IH, $ \theta_{23} $ in HO. For, $ \delta_{CP} = 257.5 ^{0}$, values of $ \theta_{13} $ around $9.0974^{0}$ to $9.1^{0}$, $9.2^{0}$to  $9.22^{0}$, $8.94^{0}$ to $8.97^{0}$, $8.82^{0}$ to $8.84^{0}$ are favoured which agrees well with the global fit value of $ \theta_{13} $ \cite{kol}. For, $ \delta_{CP} = 69 ^{0}$, values of $ \theta_{13} $ around $9.0874^{0}$ to $9.1^{0}$, $9.21^{0}$to  $9.2^{0}$, $8.945^{0}$ to $8.99^{0}$ , $8.85^{0}$  are  favoured for $5.7\times 10^{-10} < \eta_{B} < 6.7 \times 10^{-10} $ which is compatible with the global fit value of $ \theta_{13} $ \cite{kol}. For, $ \delta_{CP} = 88 ^{0}$ in Fig. 5(a), IH, $ \theta_{23} $ in LO, values of $ \theta_{13} $ around $9.0974^{0}$ to $9.103^{0}$, $9.61^{0}$to  $9.65^{0}$ are favoured for $5.7\times 10^{-10} < \eta_{B} < 6.7 \times 10^{-10} $. Similarly for, $ \delta_{CP} = 311^{0}$ in Fig. 5(e), IH, $ \theta_{23} $ in LO, values of $ \theta_{13} $ around $9.0974^{0}$ to $9.12^{0}$, $9.72^{0}$to $9.78^{0}$ are mostly favoured for $5.7\times 10^{-10} < \eta_{B} < 6.7 \times 10^{-10} $ which is consistent with the global fit data of $ \theta_{13} $ at 2$ \sigma $ and 3$ \sigma $ C.L \cite{kol}. Lastly for  $ \delta_{CP} = 95^{0}$ in Fig. 5(b), IH, $ \theta_{23} $ in HO, values of $ \theta_{13} $ around $9.0974^{0}$ to $9.11^{0}$, $9.52^{0}$to  $9.54^{0}$ are mostly favoured for $5.7\times 10^{-10} < \eta_{B} < 6.7 \times 10^{-10} $ compatible with global fitting of $ \theta_{13} $ at 2$ \sigma $ and 3$ \sigma $ C.L \cite{kol}.

\begin{table}[H]
\begin{center}
\begin{tabular}{|c|c|c|c|c|c|}
\hline 
\textbf{Case}&\textbf{hierarchy, Octant} & \textbf{$\Delta \chi^{2}$} & \textbf{$\delta_{CP}$}  & \textbf{$\epsilon_{l}$}& \textbf{$|\eta_{B}|$}\\ 
\hline 
$1$&$\text{IH, HO}$ & $4$&$ 13.5 $ & $4.91465\times10^{-6}$&$2.03389\times10^{-8} $   \\
\hline 
$2$&$\text{IH, HO}$ & $4$&$ 157.5 $ & $-7.63368\times10^{-7}$&$3.15914\times10^{-9} $ \\
\hline
$3$&$\text{IH, HO}$ & $4$&$ 202 $ & $1.24531\times10^{-6}$&$5.15362\times10^{-9} $ \\
\hline
$4$&$\text{IH, HO}$ & $4$&$ 346.3$ & $-2.09025\times10^{-9}$&$8.65034\times10^{-12} $ \\
\hline
$5$&$\text{IH, HO}$ & $9$&$ 29$ & $-5.98012\times10^{-6}$&$2.47483\times10^{-8} $\\
\hline
$6$&$\text{IH, HO}$ & $9$&$ 153 $ & $1.18773\times10^{-5}$&$4.91533\times10^{-8} $\\
 \hline
$7$&$\text{IH, HO}$ & $9$&$ 209$ &  $8.38787\times10^{-6}$ &$3.47125\times10^{-8} $\\
\hline
$8$&$\text{IH, HO}$ & $9$&$ 332.5$ & $2.45147\times10^{-7}$&$1.01449\times10^{-9} $\\
\hline
$9$&$\text{IH, LO}$ & $9$&$ 332.5$ & $1.03435\times10^{-6}$&$4.28058\times10^{-9} $   \\
\hline 
$10$&$\text{IH, LO}$ & $9$&$ 209 $ & $5.36981\times10^{-6}$&$2.22225\times10^{-8} $ \\
\hline
$11$&$\text{IH, LO}$ & $9$&$ 153 $ & $7.94367\times10^{-6}$&$3.28743\times10^{-8} $ \\
\hline
$12$&$\text{IH, LO}$ & $9$&$ 29$ & $-7.28224\times10^{-6}$&$3.0137\times10^{-8} $ \\
\hline
$13$&$\text{IH, LO}$ & $4$&$ 346.1 $ & $-1.04874\times10^{-6}$&$4.34013\times10^{-9} $\\
\hline
$14$&$\text{IH, LO}$ & $4$&$ 203 $ & $1.26601\times10^{-5}$&$5.23928\times10^{-8} $\\
 \hline
$15$&$\text{IH, LO}$ & $4$&$ 157.5$ & $-9.9942\times10^{-7}$&$4.13602\times10^{-9} $\\
\hline
$16$&$\text{IH, LO}$ & $4$&$ 13.5$ & $-3.75736\times10^{-7}$&$1.55496\times10^{-9} $\\
\hline
\end{tabular}
\end{center}
\caption{Same as in Table V, but IH is used.}
\end{table}
  
\begin{center}
\begin{figure*}[htbp]
{\centering
\begin{subfigure}[]{\includegraphics[width=8.7cm]{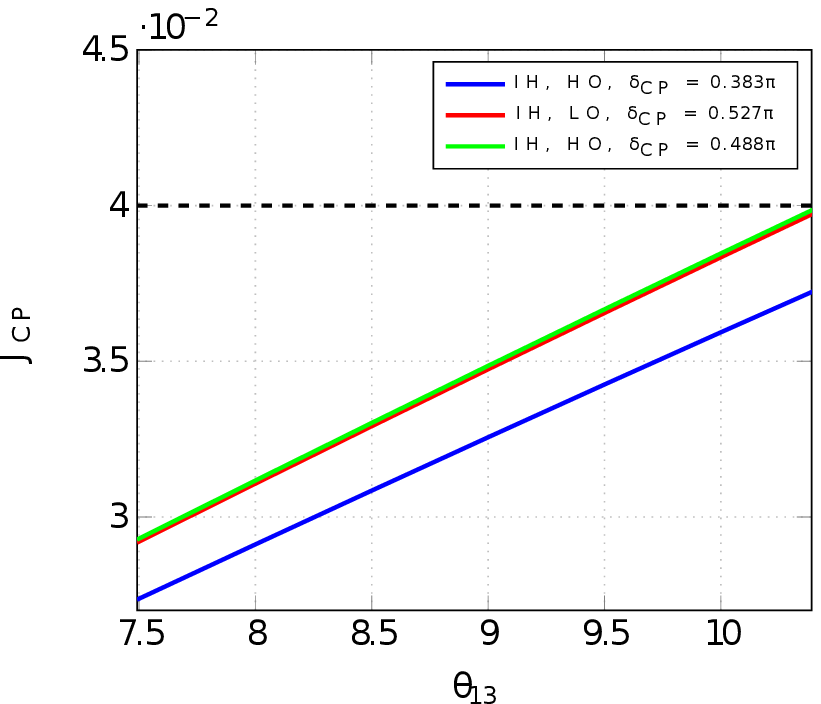}}\end{subfigure}
\begin{subfigure}[]{\includegraphics[width=8.7cm]{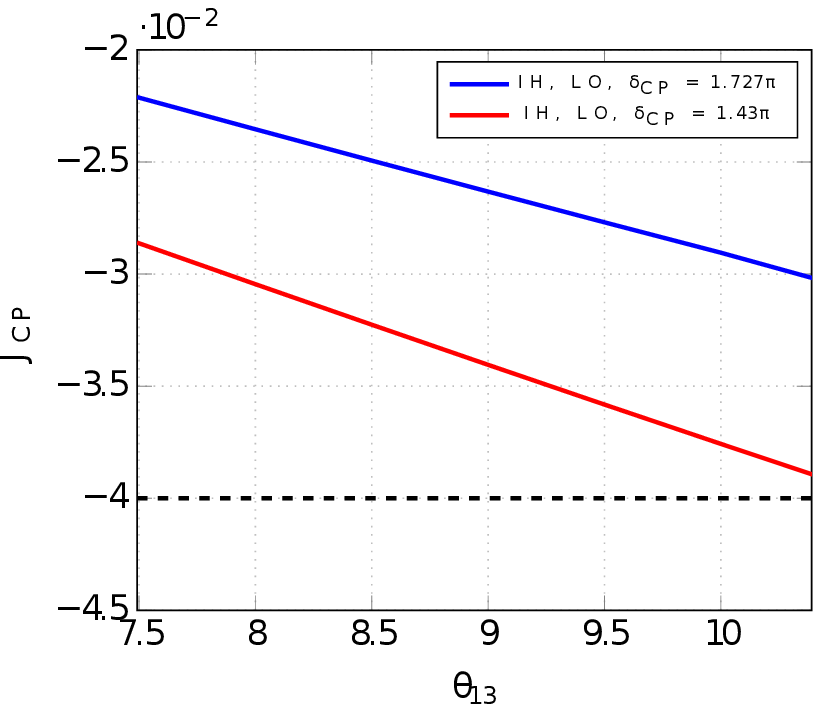}}\end{subfigure}\\
\caption{Plot of $ J_{CP} $ vrs $ \theta_{13} $ with CP phases in Fig. 6(a):$ \delta_{CP}= 95^{0} $, IH, HO; $ \delta_{CP}= 69^{0}$, IH, HO; $ \delta_{CP}= 88^{0}$, IH, LO. Fig. 6(b): $ \delta_{CP}= 257.5^{0} $, IH, HO; $ \delta_{CP}= 311^{0}$, IH, LO within the 3 $\sigma$ C.L of the best fit values of $ \theta_{13} $. Horizontal line represents the maximum allowed CP violation in the leptonic sector, $J_{CP}\leq .04|Sin\delta_{CP}|$. }}
\end{figure*}
\end{center}

The magnitude of CP violation in $ \nu_{l} \rightarrow \nu_{l^{'}} $ and $ \nu_{\bar{l}} \rightarrow \nu_{\bar{l}^{'}} $, $l=l^{'}=e,\mu,\tau$, is determined by the rephasing Jarkslog invariant $ \textit{J}_{CP} $, which in the standard parametrisation of the $ \nu $ mixing matrix has the form \cite{kol}:
\begin{equation}
\textit{J}_{CP} = Im (U_{\mu 3}U^{*}_{e3}U_{e2}U^{*}_{\mu2}) = \frac{1}{8}Cos\theta_{13}Sin2\theta_{12}Sin2\theta_{23}Sin2\theta_{13}Sin\delta_{CP}
\end{equation}
Since $Sin2\theta_{12}$, $Sin2\theta_{23}$, $Sin2\theta_{13}$ have been determined experimentally with a relatively good precision \cite{Fgli,Frero,Garc}, the size of CP violation effects in $ \nu $ oscillations depends essentially on leptonic CPV phase $ \delta_{CP} $. The current data implies $ \textit{J}_{CP} = 0.040|Sin\delta_{CP}| $ \cite{kol}, and a best fit value $\textit{J}_{CP}^{best}$ = -0.032  \cite{kol}. Our calculated values of Jarkslog invariant by plugging input for the three $ \nu $ mixing angles at its best fit for favoured cases of BAU and the values of leptonic $ \delta_{CP} $ phase are summarised in Table XIII. We find that for all the five favoured cases, our calculated values of $ J_{CP} $ lie within its present experimental limits.
\par 
In Fig. 6 we plot $ J_{CP} $ Vs $ \theta_{13} $, taking variation of $ \theta_{13} $ within 3$ \sigma $ range of its best fit value and find that the plot for all the above listed five cases, $ J_{CP} $ lies within its present experimental limits.

\begin{table}[H]
\begin{center}
\begin{tabular}{|c|c|c|c|c|c|}

\hline 
\textbf{Case}&\textbf{hierarchy, Octant} & \textbf{$\Delta \chi^{2}$} & \textbf{$\delta_{CP}$}  & \textbf{$\epsilon_{l}$}& \textbf{$|\eta_{B}|$}\\ 
\hline 
$1$&$\text{IH, HO}$ & $4$&$ 13.5 $ & $4.00427\times10^{-6}$&$1.65714\times10^{-8} $   \\
\hline 
$2$&$\text{IH, HO}$ & $4$&$ 157.5 $ & $3.11981\times10^{-6}$&$1.29111\times10^{-8} $ \\
\hline
$3$&$\text{IH, HO}$ & $4$&$ 202 $ & $3.99325\times10^{-6}$&$1.65258\times10^{-8} $ \\
\hline
$4$&$\text{IH, HO}$ & $4$&$ 346.3$ & $4.15622\times10^{-6}$&$1.72002\times10^{-8} $ \\
\hline
$5$&$\text{IH, HO}$ & $9$&$ 29$ & $4.15708\times10^{-6}$&$1.72038\times10^{-8} $\\
\hline
$6$&$\text{IH, HO}$ & $9$&$ 153 $ & $-3.99333\times10^{-6}$&$1.65267\times10^{-8} $\\
 \hline
$7$&$\text{IH, HO}$ & $9$&$ 209$ &  $-7.0083\times10^{-7}$ &$2.90033\times10^{-9} $\\
\hline
$8$&$\text{IH, HO}$ & $9$&$ 332.5$ & $-3.56253\times10^{-6}$&$1.47433\times10^{-8} $\\
\hline
$9$&$\text{IH, LO}$ & $9$&$ 332.5$ & $1.03435\times10^{-6}$&$4.28058\times10^{-9} $   \\
\hline 
$10$&$\text{IH, LO}$ & $9$&$ 209 $ & $-7.0083\times10^{-7}$&$2.90033\times10^{-9} $ \\
\hline
$11$&$\text{IH, LO}$ & $9$&$ 153 $ & $-3.99333\times10^{-6}$&$1.65261\times10^{-8} $ \\
\hline
$12$&$\text{IH, LO}$ & $9$&$ 29$ & $4.15708\times10^{-6}$&$1.72038\times10^{-8} $ \\
\hline
$13$&$\text{IH, LO}$ & $4$&$ 346.1 $ & $3.63103\times10^{-6}$&$1.50267\times10^{-8} $\\
\hline
$14$&$\text{IH, LO}$ & $4$&$ 203 $ & $-2.80629\times10^{-6}$&$1.16136\times10^{-8} $\\
 \hline
$15$&$\text{IH, LO}$ & $4$&$ 157.5$ & $3.11981\times10^{-6}$&$1.29111\times10^{-8} $\\
\hline
$16$&$\text{IH, LO}$ & $4$&$ 13.5$ & $4.000427\times10^{-6}$&$1.65714\times10^{-8} $\\
\hline
\end{tabular}
\end{center}
\caption{Same as in Table VI, but IH is used.}
\end{table}

\section{\textbf{Conclusion}}
Measuring CP violation in the lepton sector is one of the most challenging tasks today. A systematic study of the CP sensitivity of the current and upcoming LBNE/DUNE is done in our earlier work \cite{DK} which may help a precision measurement of leptonic $ \delta_{CP}$ phase. In this work, we studied how the entanglement of the quadrant of leptonic CPV phase and octant of atmospheric mixing angle $ \theta_{23} $ at LBNE/DUNE, can be broken via leptogenesis and baryogenesis. Here, we have considered the effect of ND only in LBNE, on sensitivity of CPV phase measurement, but similar conclusions would hold for the effect of reactor experiments as well. This study is done for both
 Normal hierarchy and Inverted hierarchy, Higher Octant and Lower Octant. We considered two cases of fermion rotation matrix - PMNS only, and CKM+PMNS.
Following the results of \cite{DK}, the enhancement of CPV sensitivity with respect to its quadrant is utilized here to calculate the values of lepton-antilepton symmetry. Then, this is used to calculate the  value of BAU. This is an era of precision measurements in neutrino physics. We therefore considered variation of $\Delta m^{2}_{31}$ within its 1$\sigma$, 2$\sigma$ and 3$\sigma$, and $\theta_{13}$ within its 3$ \sigma $ range. We calculated baryon to photon ratio, and compared with its experimentally known best fit value. 
\begin{table}[H]
\begin{center}
\begin{tabular}{|c|c|c|c|c|c|}

\hline 
\textbf{Case}&\textbf{hierarchy, Octant} & \textbf{$\Delta \chi^{2}$} & \textbf{$\delta_{CP}$}  & \textbf{$\epsilon_{l}$}& \textbf{$|\eta_{B}|$}\\ 
\hline 
$1$&$\text{NH, LO}$ & $16$&$ 56 $ & $1.35794\times10^{-5}$&$5.62719\times10^{-8} $   \\
\hline 
$2$&$\text{NH, LO}$ & $16$&$ 136 $ & $-3.11475\times10^{-5}$&$1.28902\times10^{-7} $ \\
\hline
$3$&$\text{NH, LO}$ & $16$&$ 232 $ & $1.37456\times10^{-5}$&$5.68851\times10^{-8} $ \\
\hline
$4$&$\text{NH, LO}$ & $16$&$ 314$ & $1.49244\times10^{-6}$&$6.17636\times10^{-9} $ \\
\hline
$5$&$\text{NH, LO}$ & $25$&$ 84 $ & $3.56574\times10^{-5}$&$1.47565\times10^{-7} $\\
\hline
$6$&$\text{NH, LO}$ & $25$&$ 122.5 $ & $7.31569\times10^{-6}$&$3.02754\times10^{-8} $\\
 \hline
$7$&$\text{NH, LO}$ & $25$&$ 263.5$ & $-1.25402\times10^{-5}$&$5.18967\times10^{-8} $\\
\hline
$8$&$\text{NH, LO}$ & $25$&$ 294.5$ & $4.28344\times10^{-6}$&$1.77267\times10^{-8} $\\
\hline
$9$&$\text{NH, HO}$ & $16$&$ 59 $ & $3.19255\times10^{-5}$&$1.32121\times10^{-7} $   \\
\hline 
$10$&$\text{NH, HO}$ & $16$&$132.5$ & $-1.70443\times10^{-5}$&$7.05367\times10^{-8} $ \\
\hline
$11$&$\text{NH, HO}$ & $16$&$ 232.25 $ & $8.92875\times10^{-6}$&$3.69509\times10^{-8} $ \\
\hline
$12$&$\text{NH, HO}$ & $16$&$ 314$ & $4.8229\times10^{-6}$&$1.99592\times10^{-8} $ \\
\hline
\end{tabular}
\end{center}
\caption{Same as in Table V, but for $ \chi^{2} = $ 16 and 25 }
\end{table}

\begin{table}[H]
\begin{center}
\begin{tabular}{|c|c|c|c|c|c|}

\hline 
\textbf{Case}&\textbf{hierarchy, Octant} & \textbf{$\Delta \chi^{2}$} & \textbf{$\delta_{CP}$}  & \textbf{$\epsilon_{l}$}& \textbf{$|\eta_{B}|$}\\ 
\hline 
$1$&$\text{NH, LO}$ & $16$&$ 56 $ & $-2.96623\times10^{-5}$&$1.2275\times10^{-7} $   \\
\hline 
$2$&$\text{NH, LO}$ & $16$&$ 136 $ & $3.22739\times10^{-5}$&$1.33563\times10^{-7} $ \\
\hline
$3$&$\text{NH, LO}$ & $16$&$ 232 $ & $-2.72203\times10^{-5}$&$1.12649\times10^{-7} $ \\
\hline
$4$&$\text{NH, LO}$ & $16$&$ 314$ & $-1.04375\times10^{-5}$&$4.31949\times10^{-8} $ \\
\hline
$5$&$\text{NH, LO}$ & $25$&$ 84 $ & $-3.32343\times10^{-5}$&$1.37537\times10^{-7} $\\
\hline
$6$&$\text{NH, LO}$ & $25$&$ 122.5 $ & $-1.47354\times10^{-6}$&$6.09812\times10^{-9} $\\
 \hline
$7$&$\text{NH, LO}$ & $25$&$ 263.5$ & $-2.36179\times10^{-5}$&$9.77404\times10^{-8} $\\
\hline
$8$&$\text{NH, LO}$ & $25$&$ 294.5$ & $-3.32892\times10^{-6}$&$1.37765\times10^{-7} $\\
\hline
$9$&$\text{NH, HO}$ & $16$&$ 59 $ & $-3.27271\times10^{-5}$&$1.35438\times10^{-7} $   \\
\hline 
$10$&$\text{NH, HO}$ & $16$&$132.5$ & $-2.97961\times10^{-5}$&$1.23309\times10^{-7} $ \\
\hline
$11$&$\text{NH, HO}$ & $16$&$ 232.25 $ & $-1.46679\times10^{-5}$&$6.07021\times10^{-8} $ \\
\hline
$12$&$\text{NH, HO}$ & $16$&$ 314$ & $-1.04375\times10^{-5}$&$4.31949\times10^{-8} $ \\
\hline
\end{tabular}
\end{center}
\caption{Same as in Table VI, but for $ \chi^{2} = $ 16 and 25 }
\end{table}

To break the quadrant of CPV phase $ - $ Octant of $ \theta_{23} $ entanglement we have calculated BAU ($ \eta_{B} $) for 152 cases as shown in Tables I-XII, and found that only for five cases, our calculated $ \eta_{B} $ lies within the present best fit values of $ \eta_{B} $. These five cases are $\delta_{CP} = 1.43\pi$ (third quadrant), $\delta_{CP} = 0.527\pi$ (second quadrant), $\delta_{CP} = .383 \pi$ (first quadrant), $\delta_{CP} = .488\pi$ (first quadrant) for the case when R matrix consists of both $V_{CKM}$ and $U_{PMNS}$ and $\delta_{CP} = 1.727\pi$ (fourth quadrant), for the case when R matrix consists of $U_{PMNS}$ only. Next, we studied variation of $ \eta_{B} $, w.r.t 1$ \sigma $, 2$ \sigma $ and 3$ \sigma $ variation of $ \Delta m^{2}_{31} $, as shown in Figs. 3 and 4. It can be seen from Fig. 3 and 4 that for variation of $\Delta m^{2}_{31}$, within its 1$\sigma$ range, all calculated values of $\eta_B$ lie in the allowed range of its best value. For $\Delta m^{2}_{31}$ at its 3 $ \sigma $ C.L, the case $ \delta_{CP} = 0.488 \pi $ is consistent with the allowed range of BAU for $ \Delta m^{2}_{31} < -2.27\times 10^{-3} eV^{2}$. Similarly, very slight discrepancy of $ \eta_{B} $ for  $ \delta_{CP} =0.5277 \pi $ can be seen from Fig. 4(a) for $ \Delta m^{2}_{31} < -2.63\times 10^{-3} eV^{2}$. Case 15 of Table XI, where $ \delta_{CP} = 1.43 \pi $ has $ |\eta_{B}| $ in the range compatible with $5.7\times 10^{-10} < \eta_{B} < 6.7 \times 10^{-10} $ except for $ \Delta m^{2}_{31} >- 2.2695 \times 10^{-3} eV^{2} $ and  $ \Delta m^{2}_{31} <- 2.635 \times 10^{-3} eV^{2} $. It is worth noting that this value of $ \frac{\delta_{CP}}{\pi}=1.43 $ is close to the central value of $ \delta_{CP} $ from the recent global fit result \cite{kol}. Case 13 of Table XI: $ \delta_{CP} = 0.383 \pi $ has $ |\eta_{B}| $ in the range allowed by, $5.7\times 10^{-10} < \eta_{B} < 6.7 \times 10^{-10} $ except for $ \Delta m^{2}_{31} >- 2.385 \times 10^{-3} eV^{2} $. Case 4 of Table XII, where $ \delta_{CP} = 1.727 \pi $, has $ |\eta_{B}| $ in the range favoured by the present experimental constraints except for $ \Delta m^{2}_{31} >- 2.255 \times 10^{-3} eV^{2} $ where the straight line fails to satisfy allowed $ |\eta_{B}| $ bounds even at $ 2 \sigma $ C.L of $ \Delta m^{2}_{31} $. Interestingly here leptonic CPV phase $ \delta_{CP} =1.727 \pi $ lies within the 1$ \sigma $ ranges of $ \delta_{CP} $ from latest global fit analysis, $ \delta_{CP} = 1.67^{+0.37}_{-0.77} $ \cite{kol}. Here  $ R_{1j} $ elements of R matrix consists of only $ U_{PMNS} $ elements.

\begin{table}[H]
\begin{center}
\begin{tabular}{|c|c|c|c|c|c|}

\hline 
\textbf{Case}&\textbf{hierarchy, Octant} & \textbf{$\Delta \chi^{2}$} & \textbf{$\delta_{CP}$}  & \textbf{$\epsilon_{l}$}& \textbf{$|\eta_{B}|$}\\ 
\hline 
$1$&$\text{IH, LO}$ & $25$&$ 59 $ & $7.35254\times10^{-6}$&$3.04279\times10^{-8} $   \\
\hline 
$2$&$\text{IH, LO}$ & $25$&$ 131.5 $ & $1.00443\times10^{-6}$&$4.15675\times10^{-9} $ \\
\hline
$3$&$\text{IH, LO}$ & $25$&$ 246.5 $ & $3.71415\times10^{-6}$&$1.53707\times10^{-8} $ \\
\hline
$4$&$\text{IH, LO}$ & $25$&$ 311$ & $3.74283\times10^{-7}$&$1.54894\times10^{-9} $ \\
\hline
$5$&$\text{IH, LO}$ & $16$&$ 42 $ & $-7.26506\times10^{-6}$&$3.00659\times10^{-8} $\\
\hline
$6$&$\text{IH, LO}$ & $16$&$ 140.5$ & $7.91014\times10^{-6}$&$3.27355\times10^{-8} $\\
 \hline
$7$&$\text{IH, LO}$ & $16$&$ 225.5$ & $8.74966\times10^{-7}$&$3.62098\times10^{-9} $\\
\hline
$8$&$\text{IH, LO}$ & $16$&$ 320.5$ & $8.74965\times10^{-7}$&$3.62097\times10^{-9} $\\
\hline
$9$&$\text{IH, HO}$ & $16$&$ 45 $ & $-1.88492\times10^{-6}$&$7.80058\times10^{-9} $   \\
\hline 
$10$&$\text{IH, HO}$ & $16$&$139$ & $-2.36693\times10^{-7}$&$9.79536\times10^{-10} $ \\
\hline
$11$&$\text{IH, HO}$ & $16$&$ 226.5 $ & $-7.81644\times10^{-7}$&$3.23477\times10^{-9} $ \\
\hline
$12$&$\text{IH, HO}$ & $16$&$ 319$ & $-3.88288\times10^{-6}$&$1.6069\times10^{-8} $ \\
\hline
$13$&$\text{IH, HO}$ & $25$&$ 72$ & $1.40342\times10^{-7}$&$5.80793\times10^{-10} $ \\
\hline
$14$&$\text{IH, HO}$ & $25$&$ 123$ & $-3.73584\times10^{-6}$&$1.54604\times10^{-8} $ \\
\hline
$15$&$\text{IH, HO}$ & $25$&$ 257.5$ & $1.48671\times10^{-7}$&$6.15262\times10^{-10} $ \\
\hline
$16$&$\text{IH, HO}$ & $25$&$ 302$ & $-7.71976\times10^{-7}$&$3.19476\times10^{-9} $ \\
\hline
\end{tabular}
\end{center}
\caption{Same as in Table IX, but IH is used.}
\end{table}

In fig. 5 we showed variations of $ \eta_{B} $ with $ \theta_{13} $, taking range of $ \theta_{13} $ within 3$ \sigma $ values of its best fit values, for the five favoured cases and find that values of $ \theta_{13} $ around $ 9.0974^{0} $ to $9.12^{0}$ (which agrees well with the current fit data \cite{kol}) are favoured as far as matching with the best fit values of $|\eta_B|$ are concerned.
\par 
We also calculated values of Jarkslog invariant $ J_{CP} $ for these five cases, and found that they lie within present experimental limits (shown in Table XIII). Variation of $ J_{CP} $ with $ \theta_{13} $, taking range of $ \theta_{13} $ within its 3$ \sigma $ values of its best fit values was also considered (Fig. 6), and find that $ J_{CP} $ lies within its experimental limits for these five cases even when variation of $ \theta_{13} $ is taken. These results could be important, as the quadrant of leptoniv CPV phase, and octant of atmospheric mixing angle $\theta_{23}$ are yet not fixed experimentally. Also, they are significant in context of precision measurements of neutrino oscillation parameters, specially the leptonic CPV phase, $ \Delta m^{2}_{31} $ and the reactor angle $\theta_{13}$.

It may be noted that out of the five cases found favourable in our work here, one of the values $ \delta_{CP} = 1.43\pi $ matches with the latest global fit value, $ \delta_{CP} $ = 1.4 $ \pi $. Future experiments like DUNE/LBNEs and Hyper-Kamionande \cite{Mth} that would measure $ \delta_{CP} $ (especially probing leptonic CPV) will support/disfavour the results presented in this work.

\section*{Acknowledgments}
GG would like to thank UGC, India, for providing RFSMS fellowship to her, during which part of this work was done. DD thanks HRI,
Allahabad, India for
providing a postdoctoral fellowship to him. KB thanks DST-SERB, Govt of India, for financial support through a project.

\begin{table}[H]
\begin{center}
\begin{tabular}{|c|c|c|c|c|c|}

\hline 
\textbf{Case}&\textbf{hierarchy, Octant} & \textbf{$\Delta \chi^{2}$} & \textbf{$\delta_{CP}$}  & \textbf{$\epsilon_{l}$}& \textbf{$|\eta_{B}|$}\\ 
\hline 
$1$&$\text{IH, LO}$ & $25$&$ 59 $ & $-4.11136\times10^{-6}$&$1.70145\times10^{-8} $   \\
\hline 
$2$&$\text{IH, LO}$ & $25$&$ 131.5 $ & $-3.26348\times10^{-6}$&$1.35057\times10^{-8} $ \\
\hline
$3$&$\text{IH, LO}$ & $25$&$ 246.5 $ & $9.54714\times10^{-7}$&$3.95101\times10^{-9} $ \\
\hline
$4$&$\text{IH, LO}$ & $25$&$ 311$ & $1.47958\times10^{-7}$&$6.12311\times10^{-10} $ \\
\hline
$5$&$\text{IH, LO}$ & $16$&$ 42 $ & $3.06981\times10^{-6}$&$1.27042\times10^{-8} $\\
\hline
$6$&$\text{IH, LO}$ & $16$&$ 140.5$ & $4.12475\times10^{-6}$&$1.707\times10^{-8} $\\
 \hline
$7$&$\text{IH, LO}$ & $16$&$ 225.5$ & $4.11818\times10^{-6}$&$1.70428\times10^{-8} $\\
\hline
$8$&$\text{IH, LO}$ & $16$&$ 320.5$ & $-4.80846\times10^{-7}$&$1.98994\times10^{-9} $\\
\hline
$9$&$\text{IH, HO}$ & $16$&$ 45 $ & $3.74039\times10^{-6}$&$1.54905\times10^{-8} $   \\
\hline 
$10$&$\text{IH, HO}$ & $16$&$139$ & $4.18492\times10^{-7}$&$1.73189\times10^{-8} $ \\
\hline
$11$&$\text{IH, HO}$ & $16$&$ 226.5 $ & $2.40081\times10^{-6}$&$9.93556\times10^{-9} $ \\
\hline
$12$&$\text{IH, HO}$ & $16$&$ 319$ & $-1.06298\times10^{-6}$&$4.39907\times10^{-9} $ \\
\hline
$13$&$\text{IH, HO}$ & $25$&$ 72$ & $-9.54837\times10^{-7}$&$3.95152\times10^{-9} $ \\
\hline
$14$&$\text{IH, HO}$ & $25$&$ 123$ & $3.41971\times10^{-6}$&$1.91552\times10^{-8} $ \\
\hline
$15$&$\text{IH, HO}$ & $25$&$ 257.5$ & $1.81927\times10^{-6}$&$7.52892\times10^{-9} $ \\
\hline
$16$&$\text{IH, HO}$ & $25$&$ 302$ & $3.04466\times10^{-6}$&$1.26001\times10^{-8} $ \\
\hline
\end{tabular}
\end{center}
\caption{Same as in Table X but IH is used.}
\end{table}

\begin{table}[H]
\begin{center}
\begin{tabular}{|c|c|c|}
\hline 
\textbf{Serial No.} & \textbf{$ \delta_{CP} $, hierarchy, octant, $ J_{CP} $ of our calculation }& \textbf{Quadrant Of $ \delta_{CP} $} \\ 
\hline 
$1. $ & $ \delta_{CP} = 1.43 \pi $, IH, HO, $\textit{J}_{CP}$ = -.03439461 & third quadrant\\ 
\hline
$2. $ & $ \delta_{CP} = 1.727 \pi $, IH, LO, $\textit{J}_{CP}$ = -.026588173 & fourth quadrant  \\
\hline
$3. $ & $ \delta_{CP} = 0.5277 \pi $, IH, HO, $\textit{J}_{CP}$ = .035095635 & second quadrant   \\
\hline 
$4. $ & $ \delta_{CP} = 0.488 \pi $, IH, LO, $\textit{J}_{CP}$ = .035208214 & first quadarnt\\
\hline
$5. $ & $ \delta_{CP} = 0.383 \pi $ , IH, HO, $\textit{J}_{CP}$ = .032889754 & first quadrant\\
\hline
\end{tabular}
\end{center}
\caption{ Preferred cases of $\delta_{CP}$, octant, hierarchy and $ \textit{J}_{CP} $ allowed by present range of  $\eta_{B}$, $5.7\times 10^{-10} < \eta_{B} < 6.7 \times 10^{-10} $  }
\end{table}

\end{document}